\documentclass[reprint,superscriptaddress,amsmath,amssymb,aps,prl]{revtex4-2}
\usepackage{graphicx}
\usepackage{dcolumn}
\usepackage{bm}
\usepackage{color}
\usepackage{units}
\usepackage{wasysym}
\usepackage{MnSymbol}
\usepackage{comment}
\usepackage{times}
\usepackage[unicode=true,pdfusetitle,bookmarks=false,colorlinks=true,citecolor=blue,urlcolor=blue,linkcolor=blue, bookmarksnumbered=true]{hyperref}
\newcommand{\mycomment}[1]{}

\usepackage{epstopdf}
\AppendGraphicsExtensions{.tif}

\begin{document}

\title{Self-assembly of anisotropic particles on curved surfaces}

\author{Gautam Bordia}
\affiliation{Department of Materials Science and Engineering, University of California, Berkeley, California 94720, USA}
\affiliation{Materials Sciences Division, Lawrence Berkeley National Laboratory, Berkeley, California 94720, USA}
\author{Thomas P. Russell}
\affiliation{Materials Sciences Division, Lawrence Berkeley National Laboratory, Berkeley, California 94720, USA}
\affiliation{Polymer Science and Engineering Department, University of Massachusetts, Amherst, Massachusetts 01003, USA}
\author{Ahmad K. Omar}
\email{aomar@berkeley.edu}
\affiliation{Department of Materials Science and Engineering, University of California, Berkeley, California 94720, USA}
\affiliation{Materials Sciences Division, Lawrence Berkeley National Laboratory, Berkeley, California 94720, USA}

\begin{abstract}

The surface curvature of membranes, interfaces, and substrates plays a crucial role in shaping the self-assembly of particles adsorbed on these surfaces.
However, little is known about the interplay between particle anisotropy and surface curvature and how they couple to alter the free energy landscape of particle assemblies.
Using molecular dynamics simulations, we investigate the effect of prescribed curvatures on a quasi-2D assembly of anisotropic patchy particles. 
By varying curvature and surface coverage, we uncover a rich geometric phase diagram, with curvature inducing ordered structures entirely absent on planar surfaces.
Large spatial domains of ordered structures can contain hidden microdomains of orientational textures imprinted by the surface on the assembly. 
The dynamical landscape is also reshaped by surface curvature, with a glass-like state emerging at modest densities and high curvature. 
Changes to the symmetry of the surface curvature are found to result in unique structures, including phases with mesoscale ordering.
Our findings show that the coupling between surface curvature and particle geometry opens a new space of morphologies and structures that can be exploited for material design.

\end{abstract}

\maketitle

The self-assembly of particles on surfaces is both ubiquitous in nature and widely realized in many synthetic systems.
Indeed, the self-assembly of amphiphillic molecules, nanoparticles, and colloids on liquid-liquid interfaces underpins nearly all emulsion-like materials~\cite{Stratford2005, Yang2022, Cates2008, Forth2021, McGlasson2024OnParticles, Shi2018} including those used in catalysis, pharmaceuticals, and in many everyday products.
In nature, the self-assembly of proteins on membrane surfaces~\cite{Cordella2014MembraneOrder, Curk2018ControllingMembranes, Yuan2021MembraneSeparation., Curtis2024DriversMembrane, Stachowiak2012MembraneCrowding, Wan2024Flower-shapedMembranes} can result in striking pattern formation and is essential for a number of biophysical processes, such as endocytosis. 
In recent years, it has become increasingly clear that the \textit{curvature} of the surface can profoundly impact the resulting self-assembled materials. 
Connections between surface geometry and structure has been identified in a number of contexts including  nematic films~\cite{Kamien2009ExtrinsicSubstrates, Matsumoto2015WrinklesCurvature, Vu2018CurvatureFilms}, colloidal crystals~\cite{Guerra2018FreezingSphere,  Tanjeem2021GeometricalNanomaterials, Sun2025ColloidalCones, Grason2016Perspective:Assemblies} and morpohogensis~\cite{Curk2018ControllingMembranes, Curtis2024DriversMembrane, Peddireddy2021Self-shapingTension, Guttman2016HowTails, Ramakrishnan2015OrganelleRemodeling, Noguchi2022MembraneDomains}.

To isolate the role of curvature, many studies have introduced simple model systems where curvature can be continuously tuned from the planar limit. 
These studies have shown that curvature can stabilize long-lived topological defects in ordered structures relative to the flat case~\cite{Vitelli2006CrystallographySurfaces, Garcia2013CrystallizationSurfaces, Garcia2015DefectCrystals, Singh2022ObservationSphere}, induce crystal misalignment between grains~\cite{Meng2014ElasticSurface, Sun2025ColloidalCones}, modify the nucleation landscape~\cite{Vitelli2006CrystallographySurfaces, Garcia2013CrystallizationSurfaces, Meng2014ElasticSurface,  Gomez2015PhaseSpace} and bias the location of condensed phases~\cite{Law2020PhaseLocation}.
The exact effect of curvature on self-assembly strongly depends on particle-level geometry and interactions.
Indeed, experimental work has shown curvature can enhance ordering in colloids with anisotropic shapes~\cite{Liu2022Curvature-assistedSurfaces} and colloids with anisotropic interactions induced by capillary forces~\cite{Shindel2012InterfacialCrystals, Ershov2013Capillarity-inducedCurvature}.
Furthermore, studies on the assembly of anisotropic particles on membranes highlight the importance of membrane-mediated particle interactions induced by surface deformations. 
In these works particle shape, density, and adhesion strength are adjusted to induce a variety of surface morphologies~\cite{Koltover1999MembraneVesicles,Olinger2016Membrane-mediatedNanoparticles,Noguchi2022MembraneDomains,Liu2023WrappingRigidity,Sharma2024HighlyVesicles}.
Intriguingly, these results suggest that the coupling of surface curvature and anisotropic particle interactions can be leveraged to induce emergent structural and dynamical features on surface assemblies.

In this work, we aim to isolate how the coupling between particle anisotropy and surface curvature influences material phase behavior and dynamics. 
We employ a ``patchy'' particle model in which the interaction anisotropy stems from the anisotropic distribution of discrete enthalpic patches (see Fig.~\ref{fig:Schematic}).
Such particles can now be realized experimentally~\cite{Wang2012ColloidsBonding, Gong2017PatchyFusion} with these directional interactions resulting in distinct phase behavior~\cite{Zhang2004, Bianchi2006,  Damasceno2012PredictiveStructures, Espinosa2019BreakdownParticles, Russo2011ReentrantFluids}. 
Here we consider the self-assembly of three-dimensional anisotropic particles confined to two-dimensional surfaces with prescribed curvature (Fig.~\ref{fig:Schematic}).
We study the phase behavior, structure, and dynamics resulting from varying curvature.
In doing so, we isolate the direct effect of surface curvature on steady-state assembly properties and uncover a rich geometric phase diagram populated with re-entrant coexistence regions in addition to glassy states. We find that curvature can profoundly impact morphology, and like temperature and pressure, can be a useful lever to modulate material properties. Our approach, which isolates the impact of static curvature on self-assembly, can constitute an important step toward the study of systems with \textit{dynamic} curvature (i.e. systems with surfaces that temporally fluctuate such as a liquid-liquid interface). 

\section{Results and Discussion}\label{sec:Results}
\subsection{Model System}  
\begin{figure}
	\centering
	\includegraphics[width=.475\textwidth]{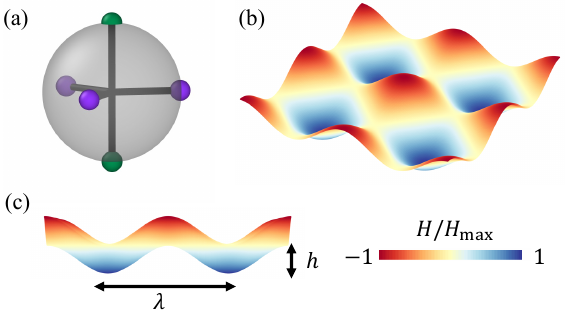}
	\caption{\protect\small{{(a) Schematic representation of a patchy particle. The core is shown in gray while the equatorial and polar patches are shown in purple and green respectively. Patches are enlarged for clarity. (b) Illustration of a subsection of the two dimensional sinusoidal surface on which the patchy particles are confined. The surface is colored by the mean curvature $H$ normalized by its maximum value $H_{\mathrm{max}}$.} (c) Side view of the surface displayed in (b).  
    }}
	\label{fig:Schematic}
\end{figure}

We model a patchy particle by considering each particle to consist of a large spherical central particle (hereafter referred to as the core particle) of diameter $d$ decorated with smaller spherical ``patch'' particles with diameter $d/10$ with each patch located at a fixed distance of $d/2$ from the center of the core.  
We fix the number of patches per particle to five and arrange patches on the particle surface to maximize the interpatch separation distances. 
For pentavalent particles, this is a trigonal planar arrangement of three equatorial patches and two polar patches [see Fig.~\ref{fig:Schematic}(a)]. 
A discussion on the effects of patch number and their spatial arrangement is provided in the Supporting Information (SI).
The relative positions of the patches and core particle are maintained through a rigid body constraint.
Patches on distinct particles interact with short-ranged attractive interactions with an energy scale of $\varepsilon_{\mathrm{patch}}=10 k_BT$ (comparable to the interaction strength of DNA coated colloids~\cite{Rogers2011DirectModeling}) while interparticle core-patch and core-core interactions are repulsive.
The forces acting on the six particles of each rigid body decompose to forces and torques acting on the center of the core particle. 
Additionally, stochastic and dissipative forces from an implicit equilibrium solvent are included.
These non-conservative forces satisfy the fluctuation-dissipation theorem, ensuring equilibrium statistics in the long-time steady state limit. 
Details for the precise functional form for all forces and the equations of motion are provided in Methods.

The three-dimensional particles are ``pinned'' to a two-dimensional surface through a strong confining harmonic potential centered at an implicitly defined surface in the Monge from: ${z - S(x,y) = 0}$. 
A particle located at position $(x_p, y_p, z_p)$ experiences a force with a magnitude proportional to $k \left( z_p - S(x_p, y_p)\right)$ where $k$ is a stiff ($k \gg 10 k_BT/d^2)$ spring constant (see Methods for implementation details).
We choose to examine surfaces represented as 2D sinusoidal waves with ${S(x,y; h, \lambda)= \frac{h}{2}\left[\cos(\frac{2\pi}{\lambda} x) +\cos(\frac{2\pi}{\lambda} y) \right]}$.  
This surface topology, which has an intrinsic square symmetry, will be the primary focus of this work. 
However, we will also later explore the effect of a surface with triangular symmetry.
The total surface area is thus ${\mathcal{A} = \iint \mathrm{d}x\mathrm{d}y \sqrt{1+(\bm{\nabla} S)^2}}$.
The two controllable curvature parameters are the amplitude ($h$) and wavelength ($\lambda$) of the wave.
Figures~\ref{fig:Schematic}(b) and (c) display the surface and indicate the spatial dependence of the mean curvature ${H(x,y)=\left(R_1^{-1} + R_2^{-1}\right)/2}$ where $R_1(x,y)$ and $R_2(x,y)$ are the local principal radii.
We choose simulation box sizes such that the total integrated mean curvature defined as ${\iint \mathrm{d}x\mathrm{d}y H\sqrt{1+(\bm{\nabla} S)^2}}$ is always zero.
All simulations were conducted with a minimum of $10000$ patchy particles using \texttt{HOOMD-blue}~\cite{Anderson2020HOOMD-blue:Simulations} with rigid body dynamics~\cite{Nguyen2011RigidUnits} and periodic boundary conditions.

Our choice of a strong spring force results in all particles irreversibly adsorbing to the surface with negligible departures from $S$ (consistent with the expected fluctuations from the equipartition theorem).
The effective binding energy of these particles is thus comparable to those found in other quasi-2D particle assemblies, including nanoparticles at liquid-liquid interfaces~\cite{Shi2018, Binks2002ParticlesDifferences}.
We do not perform simulations near or exceeding the maximal surface coverage and thus all $N$ particles in the system are bound to the surface and form a monolayer.
In addition to the dimensionless energy scale, $\varepsilon_{\rm patch}/k_BT$, the system state is described by three dimensionless geometric parameters: the surface amplitude $h/d$, the (inverse) wavelength $\mathcal{C} \equiv d/ \lambda$ and the surface coverage $\phi =  \pi d^2 N/(4\mathcal{A}$). 
We vary $\phi$ and fix the surface amplitude to a small value $h/d = 2^{-1/6}$, varying the curvature solely through $\mathcal{C}$. 
The amplitude and the range of $\mathcal{C}$ we consider are comparable to those found in natural systems such as lipid bilayer ripple phases~\cite{Kaasgaard2003Temperature-controlledBilayers, Lenz2007StructureBilayers}.

\subsection{Phase Diagram}
\begin{figure}
	\centering
	\includegraphics[width=0.475\textwidth]{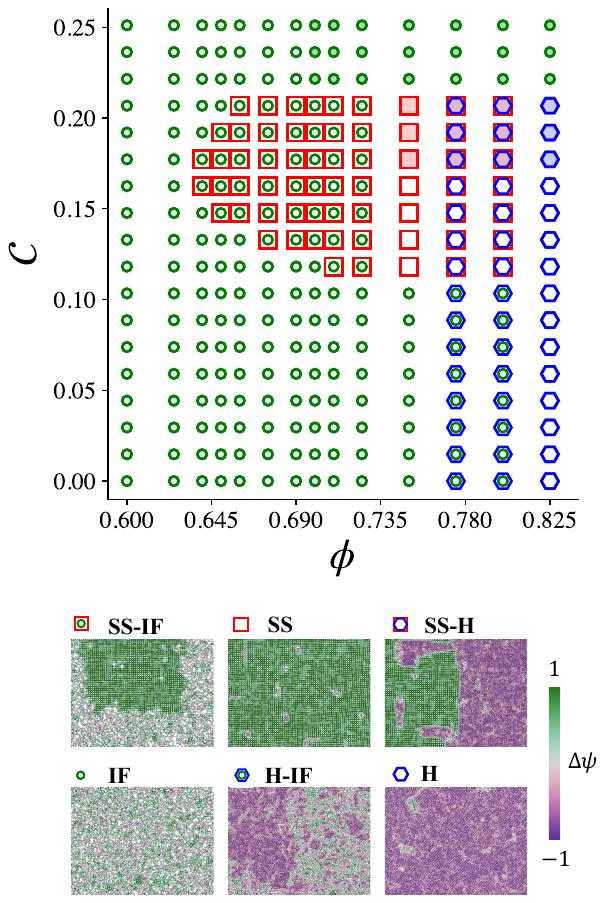}
	\caption{\protect\small{{Phase Diagram of pentavalent patchy particles confined to a 2D sinusoidal surface as a function of surface coverage $\phi$ and curvature $\mathcal{C}$, with $\varepsilon_{\mathrm{patch}} = 10 k_BT$ and $h/d = 2^{-1/6}$. The green circle morphology marker indicates the isotropic fluid (IF), the red square indicates the square solid (SS), and the blue hexagon indicates the hexagonal (H) phase. States of coexistence are marked by two symbols corresponding to the respective homogeneous phases. Shaded symbols indicate that the states observed may not be the thermodynamic ground state as the the sluggish dynamics preclude us from conclusively determining this.  Particles in the representative snapshots are colored by the difference in their two local bond orientation order parameters (details are provided in Methods) $\Delta \psi=|\psi_4| - |\psi_6|$ to distinguish tetratic (green) and hexatic (purple) order from disorder.}}}
	\label{fig:PhaseDiagram}
\end{figure}

The phase diagram for our patchy particle assembly as a function of surface curvature ($\mathcal{C}$) and coverage ($\phi$) is presented in Fig.~\ref{fig:PhaseDiagram}.
Surface curvature is found to reshape both the thermodynamics and dynamics of the assembly. 
Finite curvature can induce states of coexistence and order that are entirely absent in the planar limit while also altering the particle dynamics and even inducing a glass transition. 
Before examining each region of our reported phase diagram in detail, we first offer a brief overview of our findings.

In the planar limit ($\mathcal{C} \rightarrow 0$), the assembly is an isotropic fluid (there is no evidence of positional or orientational order) for $\phi \le 0.75$.
For larger packing fractions, a state of coexistence is observed between an isotropic fluid and a denser phase with hexagonal order.
To determine the nature of this dense phase, we performed a simulation of a system with a density slightly larger than that of the dense phase such that the system is spatially uniform. 
This homogeneous (or uniform) dense phase was determined to be a hexagonal solid by examining the spatial decay of the relevant orientational correlation function~\cite{Gasser2010MeltingDimensions, Bernard2011Two-stepTransition, Kapfer2015Two-DimensionalTransitions, Anderson2017ShapePolygons, Hajibabaei2019First-orderDisks, Li2020PhaseDimensions} (see SI for details). 
It is possible, however, that there is a hexatic to solid transition within the narrow range of densities between that of the single phase that we determined to be a hexagonal solid (at $\phi = 0.825$) and that of the coexisting dense phase (measured to have a density of $\phi \approx 0.813$). 

With increasing surface coverage from the dilute limit, 2D colloidal systems may follow several different freezing scenarios depending on the particle shape and interactions~\cite{Bernard2011Two-stepTransition, Kapfer2015Two-DimensionalTransitions, Anderson2017ShapePolygons}.
In the scenario described by Kosterlitz-Thouless-Halperin-Nelson-Young (KTHNY) theory~\cite{Kosterlitz1973OrderingSystems} a fluid continuously transitions into a $k$-atic phase with quasi-long ranged (power-law decay) $k$-fold bond orientational order (where $k$ is the kind of ordering, i.e. $k=6$ for hexatic ordering) followed by another continuous transition into a pure solid with long-ranged bond orientational order.
An alternative scenario can occur where the system exhibits a first-order fluid to solid transition with no intermediary $k$-atic phase.
In a third scenario, the fluid will first undergo a first order transition into a $k$-atic phase but will then undergo a continuous transition into a pure solid phase.
Since we observe coexistence, our system does not follow the KTHNY continuous scenario.
Determining which of the remaining two scenarios with first order transitions is observed in our simulations requires extensive computational interrogation in the small density window separating $k$-atic and solid states. 
We thus simply label these dense states with six-fold rotational symmetry as ``hexagonal'' phases and leave a detailed examination of these states and the possible curvature dependence of the hexatic-solid transition for future work. 

As we move to finite curvature, we find that curvature appears to have little influence on the assembly for $(\phi \le 0.6)$, with the assembly remaining an isotropic fluid for wavelengths as small as four particle diameters (data not shown). 
The impact of surface curvature for particles with short-ranged interactions thus appears to be greatest for moderate to high surface coverages (in this case, $\phi > 0.6)$.
For these concentrations, three regimes of curvature dependent phase behavior exist. 

At low curvature $(0 < \mathcal{C} < 0.12 )$ the system displays qualitatively similar phase behavior to the planar case.
The effect of curvature is limited to a small quantitative effect on the hexagonal-isotropic fluid coexistence boundary (see SI for a discussion on the hexagonal-isotropic coexistence boundary).
At intermediate curvature $(0.12 < \mathcal{C} < 0.21 )$ the phase diagram changes dramatically. 
We observe a phase with square order emerging from the isotropic fluid phase resulting in a square solid-fluid coexistence region, a pure square solid, and a square solid-hexagonal coexistence region depending on the precise values of $\mathcal{C}$ and $\phi$  (see Fig.~\ref{fig:PhaseDiagram}). 
At high curvature $( \mathcal{C} > 0.21 )$ the square solid and hexagonal phases are absent for all surface coverages.
Instead we observe a homogeneous disordered phase that displays distinct dynamical behavior from its low curvature counterpart.  
The notably sluggish translational dynamics at high surface coverage $(\phi > 0.7)$ indicate a glass-like phase.
These nearly arrested dynamics at large $\mathcal{C}$ and $\phi$ prevent us from making a definitive statement regarding the thermodynamic ground state of these systems. 
For these systems, we report the apparent assembly structure at the conclusion of our simulation but indicate the possible non-equilibrium nature of these states with shaded symbols in our phase diagram.
We now discuss the curvature induced transitions in detail. 

\begin{figure*}
	\centering
	\includegraphics[width=1.0\textwidth]{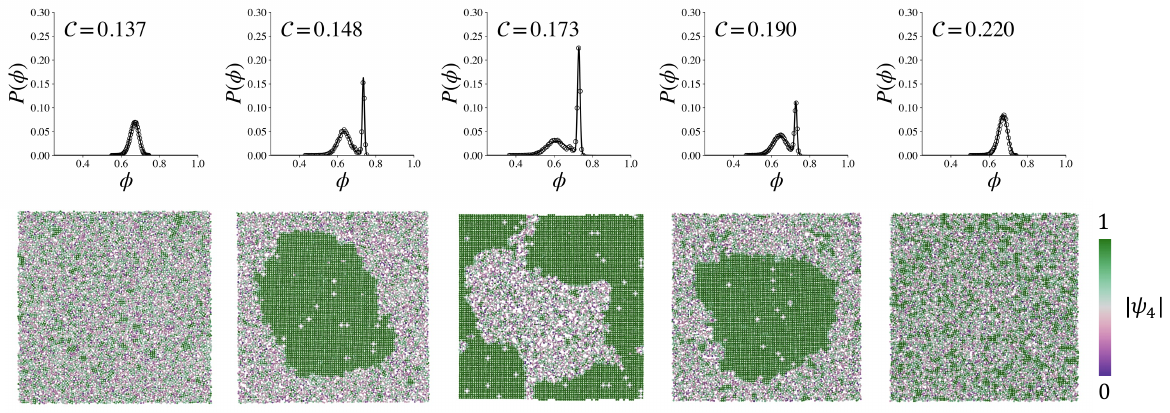}
	\caption{\protect\small{{ Local density distributions for $\mathcal{C}$ below $(\mathcal{C}= 0.137)$, inside $(0.148 \leq \mathcal{C} \leq 0.190)$ and above $(\mathcal{C}= 0.220)$ the coexistence region at bulk surface coverage $\phi = 0.668$. With increasing $\mathcal{C}$ the distribution starts as uni-modal indicating a homogeneous fluid, but then separates into two peaks at low and high surface coverage indicating phase separation. The symbols represent simulation results while solid lines indicate single or double Gaussian fits to the peaks. The height and position of the peaks are non-monotonic functions of curvature. At extreme curvature, (rightmost panel) the system returns to a disordered fluid. Simulation snapshots are included for each distribution where particles are colored by their tetratic bond orientational order parameter, $|\psi_4| $ (for details see Methods). }}}
	\label{fig:SFCoex_LocalDensityDistributions}
\end{figure*}

\subsection{Curvature induced coexistence}
In the intermediate curvature regime $(0.12 <\mathcal{C} < 0.21)$, we observe a broad range of surface coverages $(0.62<\phi<0.74)$ in which a square solid coexists with an isotropic fluid.
Within this coexistence region, we can observe the rapid nucleation and growth of solid domains with square order.
Prior theoretical work has found that surface curvature can perturb nucleation dynamics and critical nuclei size from the planar case~\cite{Horsley2018AspectsSurfaces, Gomez2015PhaseSpace}. 
For the simulations reported in Fig.~\ref{fig:PhaseDiagram}, we observe no visually discernible preferred curvature for nucleation of the solid domains. 

As the system evolves toward equilibrium, the square domains coarsen into a single solid domain that coexists with an isotropic fluid. 
In the SI, we demonstrate that for higher curvature amplitudes $h/d$, nuclei appear to only grow to a finite scale, resulting in transient fractal shapes and branching. 
These finite-size fractal domains appear similar to those observed in other context of solid assemblies on curved surfaces~\cite{Meng2014ElasticSurface, Kohler2016StressSurfaces}.
Returning to the macroscopic coexistence scenario reported in Fig.~\ref{fig:PhaseDiagram}, at steady-state, the square solid domain adopts a perimeter-minimizing shape (see Fig.~\ref{fig:SFCoex_LocalDensityDistributions} and SI for coarsening videos), indicative of a positive line tension between the coexisting domains. 

The distinct densities of the coexisting square solid and fluid are apparent in the probability distribution of the local areal density which is clearly bimodal.  
Figure~\ref{fig:SFCoex_LocalDensityDistributions} displays the surface coverage distributions as well as representative snapshots at fixed global density of $\phi=0.668$ for several values of $\mathcal{C}$ that are below, within, and above the coexistence region.
For curvatures within the coexistence region $( 0.14 <\mathcal{C} < 0.21)$, the high density peak corresponds to the solid phase, while the peak at low density corresponds to the fluid. 
We fit Gaussian distributions to each of these peaks to determine the coexisting densities of the fluid and solid phases.
By doing so for several surface coverages, we construct a square solid-fluid binodal  (see SI Fig.~S1). 
The binodal determined from this procedure engulfs the coexistence region reported in our phase diagram, indicating that some isotropic fluid states at the left boundary of our coexistence region in Fig.~\ref{fig:PhaseDiagram} are likely metastable.
As the curvature exceeds a $\phi$ dependent value, the square solid-fluid coexistence is eliminated as the system returns to a homogeneous isotropic fluid (see Fig.~\ref{fig:SFCoex_LocalDensityDistributions}).
This re-entrance to the homogeneous isotropic fluid phase occurs at progressively larger curvature values with increasing surface coverage before saturating to a curvature of $\mathcal{C} \approx 0.21$ for $\phi > 0.66$.

With the phenomenology of the (square) solid-fluid coexistence established we can now shed light on its origins. 
The driving force for square order for these finite curvatures is rooted in the coupling between local order, particle anisotropy, and curvature. 
Within the intermediate curvature region, square order facilitates increased bond formation between neighboring patches creating an energetic driving force for square order. 
At weaker curvatures, this driving force is diminished and the entropic penalty for square order precludes the formation of ordered domains. 
The elimination of square order and the return to the isotropic fluid state at the highest curvature is rooted in the changing energy landscape of the isotropic fluid. 
With increasing curvature and fixed surface coverage, bond formation is promoted in the isotropic fluid (see SI Fig.~S2) and the energetic benefit of square order is relatively diminished.
The reduced enthalpic benefit of square order relative to the fluid at these high curvatures renders the entropically favorable uniform isotropic fluid as the lower free energy state.

Surface coverages within ${0.74<\phi <0.77}$ result in a narrow region of a spatially homogeneous square solid. 
We verify that this phase is indeed a solid through the lack of spatial decay of the bond orientational correlations (see SI Fig.~S3).
Upon further increasing surface coverage to within $0.77<\phi <0.825$, we observe square solid-hexagonal coexistence.
This transition is driven by the intrinsic limitation of achieving these high surface coverages solely with square order configurations (a pure square solid cannot achieve $\phi > \pi/4$).
At these high packing fractions, a denser hexagonal phase thus emerges and coexists with the square solid to conserve global density.
We report the curvature dependence of the square solid-hexagonal coexisting densities in the the SI (see Fig.~S1). 

\begin{figure}
	\centering
	\includegraphics[width=0.475\textwidth]{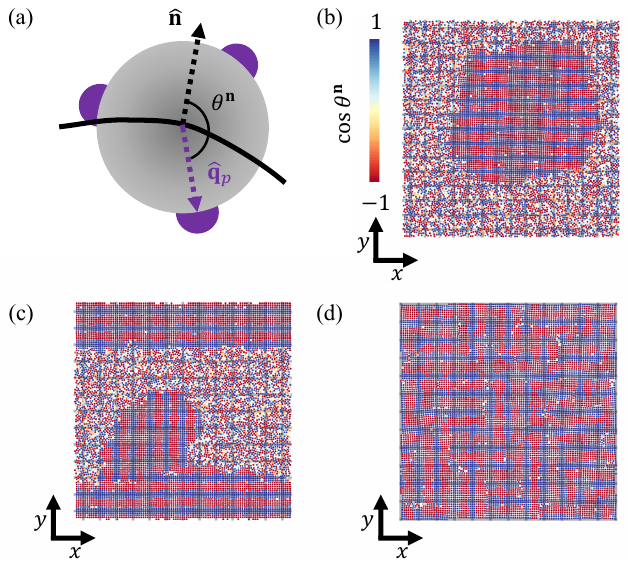}
	\caption{\protect\small{{Surface texturing of the square solid. (a) Schematic illustrating the alignment of equatorial patches (in purple) with the surface normal. Polar patches are not shown for clarity.} Orientational texturing of square solid-fluid coexistence for (b) $\phi =0.668$ and $\mathcal{C} = 0.144$; (c) $\phi =0.70$ and $\mathcal{C} = 0.126$. (d) Orientational texturing for a pure square solid, $\phi =0.75$ and $\mathcal{C} = 0.120$. For all snapshots, particles are colored by $\cos{\theta^{\mathbf{n}}}$. Anti-aligned particles are red ($\cos{\theta^{\mathbf{n}}} = -1$) and aligned particles are blue ($\cos{\theta^{\mathbf{n}}} = 1$).} To assist with visualizing the surface periodicity, we display a square grid (gray) with a spacing of $\lambda$.
    }
	\label{fig:OrientationalTexturing}
\end{figure}

\subsection{Orientational Surface Texturing}
One might expect that the periodicity of the surface wave would manifest spatially in the form of microdomains, or the presence of spatially coordinated defects in solid phases.
While some of these effects are indeed found at higher surface amplitudes (see SI), for the simulations reported here, 
the solids span many periods of the surface with no discernible effect of surface periodicity on bond-orientational order.
However, upon closely examining the positional distribution of particle orientations, we see evidence of an \textit{orientational surface texturing} that inherits the periodicity of the surface.

To maintain a square lattice structure, a patchy particle must have a specific orientation with respect to the surface to achieve four-fold patch coordination with its neighbors. 
The spatial variation in surface curvature results in a spatial variation of the particle orientations required to achieve four patch bonds.
To analyze the spatial distribution of particle orientations, we compute distributions of the angle, $\theta^{\mathbf{n}}$ between a particle's patch orientation unit vector $\hat{\mathbf{q}}_p$ and the unit surface normal $\hat{\mathbf{n}}$, shown schematically in Fig.~\ref{fig:OrientationalTexturing}(a) 
The distribution of $\theta^{\mathbf{n}}$ for a pure square solid is shown in Fig.~S5 in the SI.
We find that for each patchy particle in the square solid phase, one of its three equatorial patches is non-bonded and is instead either aligned or anti-aligned with $\hat{\mathbf{n}}$. 

Upon visualizing the spatial distribution of equatorial patch alignment with $\hat{\mathbf{n}}$ we observe a clear surface patterning.
Fig.~\ref{fig:OrientationalTexturing}(b)-(d) display representative snapshots of solid-fluid coexistence as well as a pure solid where each particle is colored by its most aligned (or anti-aligned) equatorial patch.
To display the periodic length scale of the surface, we include in the snapshots a square grid (in gray) with a spacing of $\lambda$.
The intersection of the lines forming this square grid in our snapshots represents the local maxima of the surface. 
In Fig.~\ref{fig:OrientationalTexturing}(b) we clearly observe the square solid is composed of alternating parallel stripes (of width $\lambda$) of aligned and anti-aligned domains.
Moving along the center axis of one of these lamellae coincides with moving through local maxima and saddles or through local minima and saddles. 
The selection of aligned or antialigned particle arrangements appears to be a spontaneously broken symmetry.

The length of these ``hidden lamellae'' can span the size of the square domain [Fig.~\ref{fig:OrientationalTexturing}(b)] but defects can arise that generate a ninety-degree bend in the lamellae [Fig.~\ref{fig:OrientationalTexturing}(c)].
An apparently uniform square solid can actually contain significant spatial heterogeneity in the form of these defected lamellae as shown in Fig.~\ref{fig:OrientationalTexturing}(d).
In this scenario, the defect density is quite high, with some lamellae exhibiting multiple bends over short distances. 
The thickness of an individual lamellae is never observed to deviate from the curvature wavelength, $\lambda$. 
We conclude from this analysis that despite the appearance of spatially uniformity, curvature can induce and template ``hidden'' microphases within domains. 

\begin{figure}
	\centering
	\includegraphics[width=0.475\textwidth]{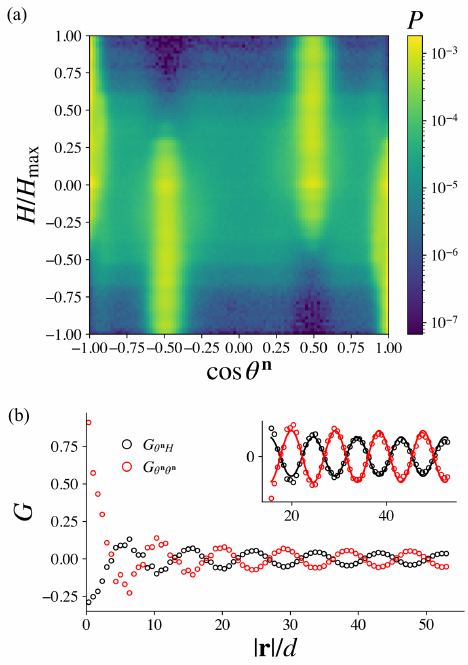}
	\caption{\protect\small{{Orientational alignment and its correlation with local curvature. (a) Joint probability distribution of the local mean curvature and the alignment of a particle's equatorial patches with the surface normal, for a pure square solid $\phi=0.75$ and $\mathcal{C}=0.120$. (b) Spatial self-correlation of particle orientations $\left(G_{\theta^{\mathbf{n}}\theta^{\mathbf{n}}}\right)$ and spatial cross-correlation between particle orientation and mean curvature $\left(G_{\theta^{\mathbf{n}}H}\right)$. Inset shows a cosine fit to the periodic undulations of each correlation function. The wavelength of these undulations correspond to $\mathcal{C}=0.120$ for both $G_{\theta^{\mathbf{n}}H}$ and $G_{\theta^{\mathbf{n}}\theta^{\mathbf{n}}}$. }
    }}
	\label{fig:OrientationDistribution}
\end{figure}

The snapshots in Fig.~\ref{fig:OrientationalTexturing} show that aligned particles are predominately located at local maxima and their connecting saddle regions.
These maxima correspond to regions of negative $H$.
To determine whether there is a quantitative coupling between local surface curvature and particle orientation in the square solid, we compute the joint probability distribution of the local mean curvature $H$ and equatorial patch alignment with the surface normal $\theta^{\mathbf{n}}$ for a pure square solid shown in Fig.~\ref{fig:OrientationDistribution}(a).
We indeed find a strong coupling between local surface curvature and particle orientation.
We observe that aligned particles ($\cos{\theta^{\mathbf{n}}} = 1$) are isolated in regions of negative mean curvature (surface maxima), while anti-aligned particles ($\cos{\theta^{\mathbf{n}}} = -1$) reside in regions of positive mean curvature (surface minima).
It can be shown through simple geometry that the intermediary peaks centered at $\cos{\theta^{\mathbf{n}}} = -0.5$ and $\cos{\theta^{\mathbf{n}}} = 0.5$ are due to the two non-aligned equatorial patches that reside on each core of aligned and anti-aligned particles respectively. 
This specific coupling between the sign of $H$ and $\theta^{\mathbf{n}}$ is observed in all domains with square order.

The joint distribution in Fig.~\ref{fig:OrientationDistribution}(a) and the periodicity of the surface suggest that there are spatial correlations between particle alignment and curvature. 
To quantify this, we compute the spatial cross correlation between the most aligned patch (equatorial patch with the largest $|\cos{\theta^{\mathbf{n}}}|$ on each core) and mean curvature ${G_{\theta^{\mathbf{n}}H}(|\mathbf{r}|) = \langle \theta^{\mathbf{n}}(\mathbf{r}) H(\mathbf{0}) \rangle}$ as well as the self-correlation between most aligned patches ${G_{\theta^{\mathbf{n}}\theta^{\mathbf{n}}}(|\mathbf{r}|)= \langle \theta^{\mathbf{n}}(\mathbf{r}) \theta^{\mathbf{n}}(\mathbf{0})\rangle}$.
Both of these exhibit periodic long-ranged spatial correlations [Fig.~\ref{fig:OrientationDistribution}(b)].
The undulations of the correlations at large distances are fit to a cosine wave whose wavelength is found to closely match the wavelength of the surface [inset of Fig.~\ref{fig:OrientationDistribution}(b)].
From this analysis, it appears that alignment of particles is directly controllable by the surface curvature and seems to indicate that both negative and positive $H$ are required for forming a square solid.

\subsection{Glassy Dynamics}
\begin{figure}
    \centering
    \includegraphics[width=.475\textwidth]{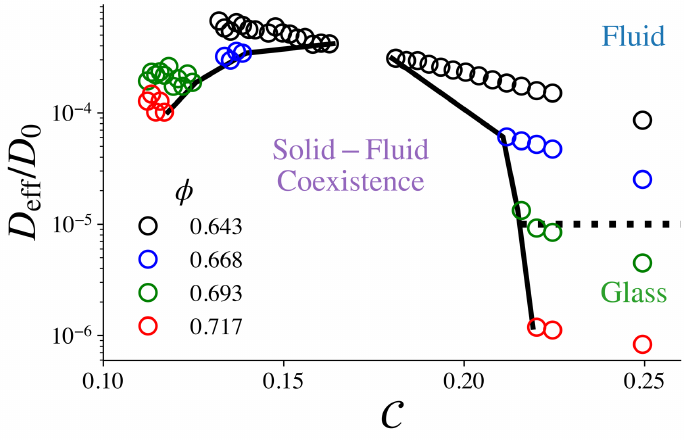}
    \caption{\protect\small{{ Diffusion coefficients as a function of curvature for several surface densities $(\phi)$ for spatially uniform states. The gap in the data indicates the presence of square solid-fluid phase separation. The diffusion coefficient is used to distinguish between the glass and fluid phases (horizontal dotted line).
    }}}
    \label{fig:DiffusionCoefficients}
\end{figure}

At curvatures above the coexistence region $(\mathcal{C}>0.21)$, the system is disordered for all surface coverages (as measured through bond-orientational correlations), but displays distinct dynamical behavior from the low curvature disordered states. 
To characterize this, we track the ensemble averaged mean-squared displacements (MSD) $\langle |\Delta\mathbf{r}(t)|^{2}\rangle $ of patchy particles, where the displacement is computed as arclengths on the surface. 
We then define an effective self-diffusion constant as measured from the MSD of patchy particles with $D_{\mathrm{eff}}= \lim_{t \to \infty} \frac{d}{dt}  \langle |\Delta\mathbf{r}(t)|^{2} \rangle/4$. 
The resulting diffusivities are shown in Fig.~\ref{fig:DiffusionCoefficients} for several densities and curvatures. 
At fixed curvature, $D_{\rm{eff}}$ decreases monotonically with surface coverage for all values of curvature.
This is consistent with our expectation and intuition from the planar limit where the increased influence of inter-particle interactions with density generally reduces mobility.  
At fixed density, we observe that $D_{\rm{eff}}$ decreases monotonically with $\mathcal{C}$ for all densities. 
At high density $(\phi>0.69)$ and curvature above the coexistence region we observe a dramatic decline in particle mobility with curvature. 
Under these conditions, we can observe diffusion constants as low as $10^{-6}D_0$ where $D_0$ is the ideal Stokes-Einstein diffusion constant in the absence of inter-particle interactions. 
We subjectively label states with self-diffusion constants below $10^{-5}D_0$ as ``glassy states'' and emphasize that these sluggish dynamics prevent us from making definitive conclusions regarding the thermodynamic ground state at these conditions.
States with these sluggish dynamics are represented through shaded symbols in Fig.~\ref{fig:PhaseDiagram}.
Notably, these immobile states display dynamic heterogeneity that can be visually appreciated by monitoring the magnitude of particle displacements with time (see SI media).

We can understand the dramatic dynamical slowing down with surface curvature by considering local particle configurations. 
Increasing $\mathcal{C}$ reduces the average Euclidean distance between particles despite the apparent surface coverage remaining constant (one can appreciate this from wrinkling a planar sheet -- while the area is conserved points drawn on the surface get closer together as the number and amplitude of wrinkles increases). 
The increasingly crowded local environment intrinsically reduces particle mobility of volume-excluding particles and additionally reduces particle motion through increased bond formation.
We expect that these considerations will hold for quasi-2D systems with both isotropic and anisotropic interactions. 

\begin{figure*}
	\centering
	\includegraphics[width=.95\textwidth]{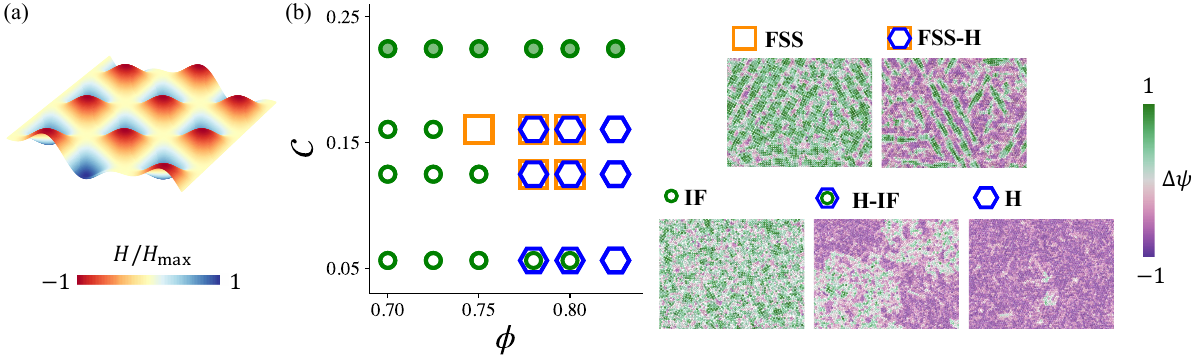}
	\caption{\protect\small{{Effect of surface symmetry on the phase behavior of pentavalent patchy particles. (a) Illustration of a subsection of the two dimensional sinusoidal surface with triangular symmetry. The surface is colored by the mean curvature. (b) Phase diagram of pentavalent patchy particles as a function of surface coverage $\phi$ and curvature $\mathcal{C}$. The morphology markers are the same as described in Fig.~\ref{fig:PhaseDiagram}, except the square solid and square solid-hexagonal coexistence morphologies are replaced by frustrated square solid (FSS) and frustrated square solid-hexagonal coexistence (FSS-H). Particles in the representative snapshots are colored by the difference in their two local bond orientation order parameters, $\Delta \psi = |\psi_4| - |\psi_6|$ to distinguish tetratic (green) and hexatic (purple) order from disorder.} }}
	\label{fig:TriangularWavePhaseDiagram}
\end{figure*}

\subsection{Effect of Surface Symmetry}
The symmetry of the surface $S(x,y)$ may play a nuanced role in determining the stability of the phases we have reported. 
So far, our choice of $S(x,y)$ has been restricted to a surface with intrinsic square symmetry which matches the inherent symmetry of the square solid phase that emerged as a result of finite curvature.
A symmetry mismatch between the surface and curvature induced phases may impact the stability of these phases.
To investigate this, we conduct simulations on a triangular surface [see methods for the precise form of $S(x,y)$]. 
A schematic of this triangular surface is shown in Fig.~\ref{fig:TriangularWavePhaseDiagram}(a).
To compare our results to the phase diagram for a square symmetric surface (Fig.~\ref{fig:PhaseDiagram}), we fix the amplitude to $h/d = 2^{-1/6}$ and choose values for $\phi $ and $\mathcal{C}$ where we expect to observe all six morphologies present in the phase diagram for the square wave. 
The results from our simulations are summarized in Fig.~\ref{fig:TriangularWavePhaseDiagram}(b). 

At low curvature ($\mathcal{C}<0.06)$ we recover the same phases as observed on the square symmetric surface which, in order of increasing surface coverage, are isotropic fluid, hexagonal-isotropic fluid coexistence and pure hexagonal.
However, for intermediate curvatures ($0.06 < \mathcal{C} < 0.21$) we observe several notable qualitative differences in phase behavior depending on the surface symmetry.
Square solid-isotropic fluid coexistence is completely absent for the triangular surface.
In its place, we observe isotropic fluid states.
At surface coverages and curvatures where we observed a pure square solid phase on a square symmetric surface (e.g.,~$\phi =0.75$, $\mathcal{C} = 0.16$) we instead observe an assembly of finite-sized square solid clusters that do not coarsen into a single large domain. 
We label these state as ``frustrated square solids'' in Fig.~\ref{fig:TriangularWavePhaseDiagram}(b).
Additionally, at surface coverages and curvatures where we observed a square solid coexisting with a hexagonal phase for the square surface (e.g.,~$\phi =0.8$, $\mathcal{C} = 0.16$), we instead see ``frustrated square solid''-hexagonal coexistence.
At high curvature ($\mathcal{C} > 0.21$) both hexagonal and square phases are absent at all surface coverages. 
As in the phase diagram for the square wave, we observe a homogeneous disordered phase that displays glass-like dynamics at the highest surface curvatures.

These results indicate that the formation of large square domains is severely hindered by the change in surface symmetry from square to triangular. 
This perhaps should be expected, as the square solid has intrinsic square symmetry while square order is a mismatch with the triangular substrate.
This mismatch likely disrupts long-range square order. 
In the SI we provide further analysis on the effect of surface geometry on the characteristics of the mesoscale domains with square order. 
We find that the long-axis of the square-ordered clusters align with the edges of the unit cell of the triangular surface (see Fig.~S8). 
Upon measuring the joint distribution of orientational alignment and mean curvature for frustrated square solid states, we find similar correlations between curvature and particle orientation (see Fig.~S9) as for the square domains on the square symmetric surface. 
Thus, the orientational texturing mechanism driving the formation of square order is preserved upon switching surface symmetry.
From these results we can conclude that the surface symmetry has a strong impact on the square solid phase and in the case of triangular surface, long-ranged square order is inhibited by the symmetry mismatch.

\subsection{Probing Curvature Dependent Thermodynamics}
\begin{figure}
	\centering
	\includegraphics[width=0.475\textwidth]{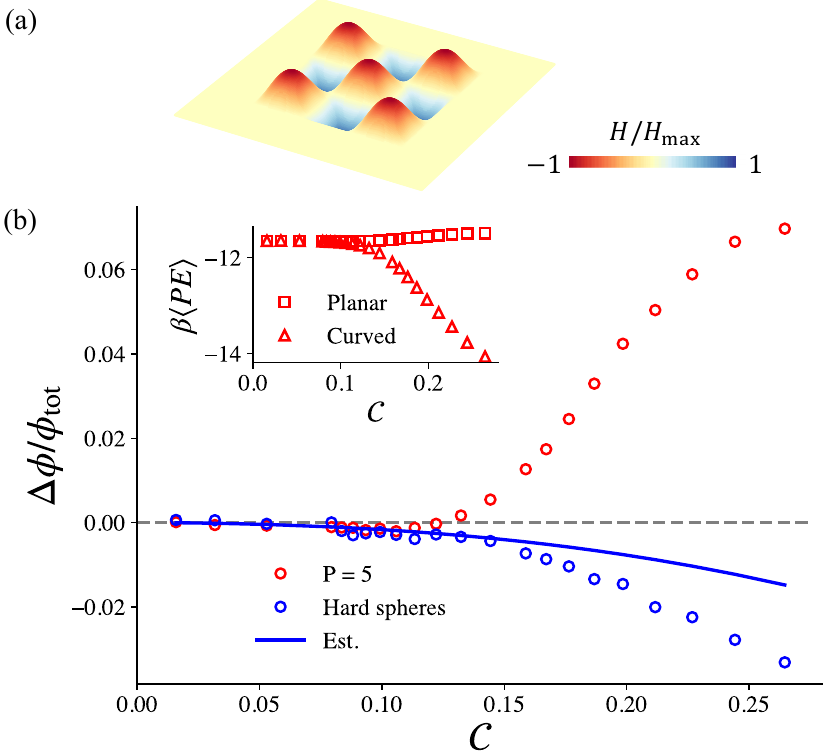}
	\caption{\protect\small{{ Probing curvature thermodynamics. (a) Schematic of the surface, which contains a curved region in the center surrounded by a planar region. The surface is colored by the mean curvature $H$. (b) Difference in surface coverage between the curved and planar regions, $\Delta \phi$, as a function of curvature $\mathcal{C}$, at a fixed global density of $\phi_\mathrm{tot}=0.6$. The red markers are for pentavalent patchy particles and the blue markers are for hard-sphere particles. The solid blue line is our prediction using a scaled particle theory (SPT) equation of state for the excess chemical potential that accounts for the change in per-particle excluded volume as a function of $\mathcal{C}$. The inset shows the per-particle potential energy for patchy particles in the curved and planar region. }
    }}
	\label{fig:DeltaPhi}
\end{figure}

Theories of phase behavior require an understanding of the free energy landscape. 
More specifically, we could build a complete thermodynamic description of the phase behavior presented in this study if we had an equation of state for the chemical potential of our particles.
Crucially, this chemical potential will now also be a \textit{function of the surface curvature}, $\mathcal{C}$.
To probe how curvature effects the chemical potential, we devise a system in which part of the surface is flat while the remaining surface is curved (square periodicity) with curvature $\mathcal{C}$ [see Fig.~\ref{fig:DeltaPhi}(a)].
The curved region is at the center and its boundaries are chosen such that the surface area of the curved and planar regions are equivalent (see SI for implementation details). 
We can then conduct a simulation and monitor if particles migrate to or away from the curved region, reporting the equilibrium density difference between the two regions, $\Delta \phi = \phi_{\mathcal{C}}-\phi_{\mathcal{P}}$, where $\mathcal{C}$ denotes the curved region and $\mathcal{P}$ denotes the planar region.
A positive $\Delta \phi$ would suggest particles have a ``preference'' for the curved region, and a negative $\Delta \phi$ the converse.
Moreover, as the chemical potential must be spatially uniform at equilibrium the particles in the two regions must have the same chemical potential. 
Chemical potential equality results in $\mu(\phi_\mathcal{C},\mathcal{C}) = \mu(\phi_\mathcal{P},0)$ where we now consider the chemical potential to be a function of both the surface coverage and curvature. 
As thermodynamic stability requires $\partial \mu / \partial \phi > 0$, measurements of $\Delta \phi > 0$ may suggest $\partial \mu/\partial{C}<0$ for small $\mathcal{C}$ (this is formally shown in the SI). 
We can thus indirectly probe the thermodynamics of curved surfaces with this methodology. 

Figure~\ref{fig:DeltaPhi} shows our results for $\Delta \phi$ at several values of $\mathcal{C}$ at a fixed global surface coverage of ${\phi_{\mathrm{tot}}=0.6}$ for hard-sphere particles and pentavalent patchy particles. 
We choose ${\phi_{\mathrm{tot}}=0.6}$ so that no square solid nucleation occurs for the entire curvature range ${0 <\mathcal{C} < 0.28}$.
At low curvature ($\mathcal{C}<0.08$) we see for both patchy particles and hard-sphere particles there is a negligible difference in surface coverage between the curved and planar region, suggesting the chemical potential is largely unaffected by surface curvature.

At intermediate curvature ($0.08 < \mathcal{C} <0.11$) there is a slight bias toward the planar region for both patchy and hard-sphere particles, indicating that $\partial \mu / \partial \mathcal{C} > 0$. 
Here, excluded volume effects appear to be more significant with increasing curvature, raising the chemical potential. 
We can in fact understand this quantitatively by extending the scaled particle theory (SPT) equation of state for the chemical potential of hard spheres to curved surfaces (see SI for derivation) following Ref.~\cite{Lishchuk2009}. 
At intermediate curvature, the prediction from SPT matches our simulation results for both patchy particles and hard spheres, and continues to match the simulation results for hard spheres until a curvature of $\mathcal{C} \approx 0.15$.
Thus, within this range of $\mathcal{C}$, the surface coverage difference is attributed to the additional entropic penalties incurred by particles in the curved region, resulting in $\partial \mu / \partial \mathcal{C} > 0$.

At high curvature ($\mathcal{C}>0.11$), $\Delta \phi$ increases dramatically with $\mathcal{C}$ for patchy particles and even changes sign. 
The sign change indicates that $\partial \mu/\partial \mathcal{C}<0$. 
In contrast, for hard-sphere particles, we continue to see $\Delta \phi$ decrease with $\mathcal{C}$ for all surface curvatures. 
The prediction of SPT fails at the largest surface curvatures where its approximations are expected to no longer hold (see SI).
Since the entropic contribution to the chemical potential for patchy particles increases with increasing curvature (which would indicate $\partial \mu/\partial \mathcal{C}>0$), the increase of $\Delta \phi$ for patchy particles is attributed to an increasing \textit{enthalpic} drive toward populating the curved region. 
Indeed, upon computing the per-particle potential energy of patchy particles in the curved and planar region as a function of $\mathcal{C}$ [inset of Fig~\ref{fig:DeltaPhi} (b)], we find a monotonic decrease within the curved region, due to additional favorable patch-patch interactions that are enabled by curvature.
This ultimately results in $\partial \mu / \partial \mathcal{C}<0$ for $\mathcal{C}>0.11$.
The potential energy of particles in the planar region gradually increases due to a decreasing $\phi_{\mathcal{P}}$ with $\mathcal{C}$, resulting in fewer attractive interactions per particle (on average).
These results provide some preliminary insight into the effects of curvature on assembly thermodynamics. 
Intriguingly, our results suggest curvature impacts both the enthalpic (through patch-patch interactions) and entropic (through excluded volume) contributions fo the chemical potential. 

\section{Conclusions}\label{sec:conclusions}

In this work we explored the effects of curvature on self-assembly using a model system of 3D patchy particles confined to a periodic 2D wave.
We found that by adjusting curvature ($\mathcal{C}$) and surface density ($\phi$) one can induce phase separation, orientational surface texturing, and glassy dynamics.
Systematically exploring the $\mathcal{C}$-$\phi$ space uncovers a rich geometric phase diagram populated by six distinct morphologies: pure fluid, square solid-isotropic fluid coexistence, pure square solid, hexagonal-isotropic fluid coexistence, square solid-hexagonal coexistence, and pure hexagonal, as summarized in Fig.~\ref{fig:PhaseDiagram}.
Hidden within states of coexistence are orientational textures imprinted by the surface on square ordered domains.
Surface curvature also strongly impacts particle mobility and can even lead to the emergence of dynamical arrest. 
These structural and dynamical changes arise due to the changing energetic and entropic landscape induced by curvature, including the promotion of particle configurations with increased bond formation and the increased role of volume exclusion for highly curved surfaces.
These considerations lead to both the emergence and elimination of square order with $\mathcal{C}$ and states of coexistence entirely absent in the planar limit. 
Thus, we find surface curvature controls the stability of the square solid relative to the fluid.

Prior work~\cite{Vitelli2006CrystallographySurfaces, Garcia2013CrystallizationSurfaces, Garcia2015DefectCrystals, Singh2022ObservationSphere, Meng2014ElasticSurface, Sun2025ColloidalCones} has focused on the disordering effect of curvature for ordered assemblies. 
In our system, curvature plays a number of roles: it couples strongly to particle geometry and interactions to cause translational and orientational transitions that include both disorder-to-order and order-to-disorder transitions with increasing $\mathcal{C}$.
The free energy landscape of particle assembly on 2D surface can be altered in intriguing an unexpected ways, suggesting that surface curvature can be a useful axis for developing innovative engineered materials with controlled morphologies.
It is our hope that by mapping a phase diagram for a model system and highlighting the unique morphological and dynamical transitions imparted by varying curvature, we inspire experiments to further explore the role of surface geometry in 2D assemblies.

Studies on the assembly of anisotropic particles on membranes highlight the importance of membrane-mediated interactions induced by surface deformations. 
In these works particle shape, density, and adhesion strength are adjusted to induce a variety of surface morphologies~\cite{Koltover1999MembraneVesicles,Olinger2016Membrane-mediatedNanoparticles,Noguchi2022MembraneDomains,Liu2023WrappingRigidity,Sharma2024HighlyVesicles}.
Our discussion in the present study has been limited to self-assembly on static surfaces and did not consider the effects of a responsive dynamic surface that itself evolves with the particle assembly. 
In future studies, we aim to more intimately investigate the connection between surface deformations and self-assembly of patchy particles. 
Nevertheless, the sensitivity of surface coverage to curvature discussed in this work suggests that patchy particles assemblies can have strong curvature preferences, perhaps not dissimilar from those of protein assemblies on lipid membranes where surface curvature induces aggregation~\cite{Curtis2024DriversMembrane}.
We hope that the present work will motivate future studies to further explore these possible connections.

\section{Methods}\label{sec:methods}
Each of the $N$ patchy particles experiences four forces: a conservative pairwise interparticle force $\mathbf{F^{\rm int}}[\mathbf{r}^N]$ where $\mathbf{r}^N$ is the set of all core particle positions, a surface confining force $\mathbf{F^{\rm surf}}[\mathbf{r}]$, a drag force $\mathbf{F}^{\rm{drag}}=-\zeta^T \dot{\mathbf{r}}$, proportional to the particle velocity $\dot{\mathbf{r}}$ with translational drag coefficient $\zeta^T$, and a Gaussian random force $\mathbf{F}^{\rm{B}}$, which has mean $\bm{0}$ and variance ${\langle \mathbf{F}^{\rm{B}}(t)\mathbf{F}^{\rm{B}}(t')\rangle = 2\zeta^{T}k_BT\mathbf{I}\delta(t-t')}$.  
In addition, each patchy particle experiences three torques: a conservative torque derived from interparticle interactions $\bm{\tau}^{\rm int} = \sum_{p}\left(\mathbf{r}_p - \mathbf{r}\right)\times \mathbf{F}_{p}$, where $\mathbf{r}_p$ is the position of patch $p$ and $\mathbf{F}_p$ is the net interaction force felt by that patch, a rotational drag torque $\bm{\tau}^{\rm drag}=-\zeta^{R}\bm{\omega}$, where $\bm{\omega}$ is the angular velocity and $\zeta^R$ the rotational drag coefficient, and stochastic diffusive rotary torque $\bm{\tau}^{\rm B}$ with mean $\mathbf{0}$ and variance $\langle \bm{\tau}^{\rm{B}}(t)\bm{\tau}^{\rm{B}}(t')\rangle = 2\zeta^{R}k_BT\mathbf{I}\delta(t-t')$. 

The underdamped Langevin equations for translational and angular momentum follow as 
\begin{eqnarray}
\dot{\mathbf{p}} &=& \mathbf{F}^{\rm int}  + \mathbf{F}^{\rm surf} + \mathbf{F}^{\rm drag} +\mathbf{F}^{\rm B} \label{translational},\\
I \dot{\pmb{\omega}} &=& \bm{\tau}^{\rm int} + \bm{\tau}^{\rm{drag}} + \bm{\tau}^{\rm B}, \label{rotational}
\end{eqnarray}
where $I$ is the scalar moment of inertia as we approximate the moment of inertia tensor as isotropic for our nearly spherical particles.

Interparticle forces for core-core and core-patch interactions result from a Weeks-Chandler-Anderson (WCA) potential $\mathbf{F}^{\rm{int}}[\mathbf{r}; \overline{\sigma}, \varepsilon ] = -\bm{\nabla} u^{\rm{WCA}}[r; \overline{\sigma}, \varepsilon]$~\cite{Weeks1971RoleLiquids} with:
\begin{equation}
\label{eq:wca}
u^{\rm{WCA}}(r; \overline{\sigma}, \varepsilon) =
 \begin{cases}
   4\varepsilon\left[\left(\frac{\overline{\sigma}}{r}\right)^{12} - \left(\frac{\overline{\sigma}}{r}\right)^{6}\right], & r \le 2^{1/6}\overline{\sigma} \\
    0, & r > 2^{1/6}\overline{\sigma},  \\
    \end{cases}
\end{equation}
where $\overline{\sigma}$ is the arithmetic mean of the Lennard-Jones diameters of the two interacting particles and $r$ is magnitude of the separation distance. 
We set the Lennard-Jones diameter of the core and patch particles to $\sigma$ and $\sigma/10$ respectively. 
Despite using a continuous potential, hard-sphere statistics are effectively approximated by choosing a sufficient stiffness potential depth of $\varepsilon = 100 k_BT$.
For the patch-patch interactions we use a standard 12-6 Lennard-Jones potential (i.e.,~Eq.~\eqref{eq:wca} with a cutoff distance of $2.5(\sigma/10)$ and a patch interaction energy of $\varepsilon_{\rm patch}$).  
We define the core particle diameter as $d = 2^{1/6}\sigma$ (the interparticle distance in which particles exclude volume) and use the diameter as the characteristic particle scale. 

Each core particle is ``pinned'' to the surface $S$ by $\mathbf{F}^{\rm{surf}}[\mathbf{r}] = - \bm{\nabla}u^{\rm{surf}}$ where $u^{\rm{surf}}$ is a scalar potential with the following harmonic form:

\begin{equation}
u^\text{\rm{surf}}(\mathbf{r}; k, h, \lambda ) = \frac{k}{2} \bigg(z- S(x,y; h, \lambda) \bigg)^2.
\end{equation}
The resulting confining force follows as ${\mathbf{F}^{\rm{surf}}[\mathbf{r}]= -k(z - S(x,y))\mathbf{n}}$ where ${\mathbf{n} = \bm{\nabla}(z- S(x,y))}$ is normal to the surface $S(x,y)$.
The exact functional form of $S$ for the square symmetric surface is described in the main text. 
For each $\mathcal{C}$ we adjust the length of our square simulation box $L$ so that it is always an integer multiple of $\lambda$ ensuring that when a particle crosses the periodic boundaries of the box it smoothly moves along $S$.

The functional form of the triangular surface is: 
\begin{widetext}
    \begin{equation}
    \begin{split}
    S(x,y) = \frac{2\sqrt{3}h}{9}\left[\sin\left(\frac{2\pi}{\lambda}\left(\frac{\sqrt{3}}{2} x + \frac{1}{2}y\right)\right)  +\sin\left(\frac{2\pi}{\lambda}y \right) + \sin\left(\frac{2\pi}{\lambda}\left(\frac{\sqrt{3}}{2} x - \frac{1}{2}y\right)\right)   \right].
    \end{split}
    \end{equation}
\end{widetext}
For simulations on the triangular surface we use a triclinic simulation box. 
To ensure smooth periodic boundaries the angle between the edges of the box in the $x$ and $y$ directions are chosen to match the unit cell of the triangular surface and the lengths of the edges are chosen so that an integer number of unit cells comprise the surface for each $\mathcal{C}$.

In our simulations, we define the unit time as ${\tau= \zeta^T\sigma^2/k_BT}$ and integrate our equations-of-motion with a timestep $10^{-3}\tau$.
Each trajectory is first equilibrated over a duration of $10^3 \tau$ while slowly increasing $k$ up to ${k=500 k_BT/\sigma^2}$ and then simulating for a a minimum of $2\times 10^6 \tau $ to reach steady-state. 
Simulations are conducted with fully three-dimensional orientational and translational equations of motion, but due to the translational confinement of $\mathbf{F}^{\rm{surf}}$, we effectively sample a quasi-2D assembly.

To quantify different forms of order in our system we first project particle positions onto the two dimensional $x-y$ plane and compute the $k$-atic bond-orientational order parameter (using the \texttt{freud} library~\cite{Ramasubramani2020Freud:Data}):
\begin{equation}
\psi_{k,p} = \frac{1}{k} \sum_j^k \exp{(\mathrm{i}k\theta_{pj})},
\end{equation}
where the sum is over the $k$-nearest neighbors of particle $p$ and $\theta_{p,j}$ is the angle between the vector (projected onto the $x-y$ plane) connecting the center of particle $p$ with that of its neighbor $j$ and an arbitrary vector of fixed direction in the $x-y$ plane. 
A projection onto the $x-y$ plane is suitable for surfaces with low overall curvature such as the those considered in this work.
Depending on the value of $k$, this order parameter quantifies distinct forms of ordering.
For example, particles with $|\psi_4| = 1$ have \textit{tetratic} order and are locally in a square lattice configuration with 4-fold coordination while particles with $|\psi_6|=1$ have \textit{hexatic} order and are arranged in a hexagonal lattice with 6-fold coordination.
Particles in the fluid phase will have, on average,  $|\psi_4| = |\psi_6| = 0$, but may instantaneously posses high 4-fold or 6-fold symmetry due to transient arrangements of high order.
To distinguish between tetratic and hexatic ordering simultaneously, we define an order parameter $\Delta \psi = |\psi_4| - |\psi_6|$, which takes values between $[-1,1]$. 
The snapshots in Fig.~\ref{fig:PhaseDiagram} are colored by $\Delta \psi $.

To compute the local density distributions in Fig.~\ref{fig:SFCoex_LocalDensityDistributions} we first segment the surface $S(x,y)$ into a square grid and replace particle positions with a Gaussian blur to obtain a coarse-grained approximation of the surface coverage~\cite{Ramasubramani2020Freud:Data}.
The distribution of $\phi$ displayed in Fig.~\ref{fig:SFCoex_LocalDensityDistributions} represents the distribution of the local density within our grid cells.
The location of the peaks in the distribution of $\phi$ provides the density of the isotropic fluid, square and hexagonal phases for states of homogeneity or phase separation. 
The location of peaks are used to construct the coexistence boundaries (see SI).

\section{acknowledgments}\label{sec:acknowldgements}
We thank David King for helpful discussions. 
We also thank the anonymous reviewers for their helpful comments and suggestions, which improved the clarity and quality of this manuscript.
This work is supported by the U.S. Department of Energy, Office of Science, Office of Basic Energy Sciences, Materials Sciences and Engineering Division under Contract No. DE-AC02-05-CH11231 within the Adaptive Interfacial Assemblies Towards Structuring Liquids program (KCTR16). 
G.B. acknowledges partial support from the National Defense Science and Engineering Graduate fellowship.
This research used the Savio computational cluster resource provided by the Berkeley Research Computing program. 
The data that support the findings of this study are available from the corresponding author upon reasonable request.

\section{Supporting Information}\label{sec:supportinginformation}

See Supporting Information for videos illustrating various points on the phase diagram and distinct dynamical behavior and discussion regarding constructing the phase diagram, implications of particle geometry, orientational correlation functions, alignment distributions, effect of curvature on bond formation, analysis on the chemical potential of particles on a curved surface, and further details on the effect of surface symmetry.

\end{document}


\title{Supporting Information -- Self-assembly of anisotropic particles on curved surfaces}

\author{Gautam Bordia}
\affiliation{Department of Materials Science and Engineering, University of California, Berkeley, California 94720, USA}
\affiliation{Materials Sciences Division, Lawrence Berkeley National Laboratory, Berkeley, California 94720, USA}
\author{Thomas P. Russell}
\affiliation{Materials Sciences Division, Lawrence Berkeley National Laboratory, Berkeley, California 94720, USA}
\affiliation{Polymer Science and Engineering Department, University of Massachusetts, Amherst 01003, USA}
\author{Ahmad K. Omar}
\email{aomar@berkeley.edu}
\affiliation{Department of Materials Science and Engineering, University of California, Berkeley, California 94720, USA}
\affiliation{Materials Sciences Division, Lawrence Berkeley National Laboratory, Berkeley, California 94720, USA}

\maketitle

\section{Supplemental Videos}
The videos included in the Supporting Information are intended to serve as representative examples of the states/transitions listed below. For reference to the expected globally stable states as a function of ($\mathcal{C}$, $\phi$), see the phase diagram provided in the main text [Fig.~2]. In all videos, the captioned time is in units of $ \zeta^T\sigma^2/k_BT$. All videos are available at: \url{https://berkeley.box.com/s/xvzjx3yv4y0a72tqkri8jh8jb9uh99hn}.
\begin{enumerate}
    \item \textbf{Square solid nucleation} \newline squareSolid\_nucleation.mp4 \newline ($\mathcal{C} = 0.165$, $\phi=0.643$) \newline Illustrates nucleation and growth of square crystallites from the fluid phase.
    \item \textbf{Square solid-isotropic fluid coexistence } \newline squareSolid\_fluid\_coexistence.mp4 \newline ($\mathcal{C} = 0.144$, $\phi=0.668$) \newline Steady-state dynamics of a square solid showing characteristic capillary fluctuations at the solid-fluid interface.
    \item \textbf{Square nucleation at high curvature } \newline squareSolid\_nucleation\_highCurvature.mp4 \newline ($\mathcal{C} = 0.222$, $\phi=0.693$, $h/d = 2^{-7/6}$) \newline Frustrated nucleation of square solid crystallites from the fluid phase at high curvature. 
    \item \textbf{Orientational texturing of a square solid} \newline squareSolid\_fluid\_orientational\_texturing.mp4 \newline ($\mathcal{C}= 0.144$, $\phi=0.668$) \newline Orientational texture of a Square solid showing a single orientational grain. 
    \item \textbf{Square solid-hexagonal coexistence} \newline squareSolid\_hexagonal\_coexistence.mp4 \newline ($\mathcal{C} = 0.128$, $\phi=0.8$) \newline Coexistence of the square solid and hexagonal phases.
    \item \textbf{Fluid phase Mobility } \newline fluid\_mobility.mp4 \newline ($\mathcal{C} = 0.110$, $\phi=0.693$) \newline Mobility of the isotropic fluid phase.
    \item \textbf{Glass phase mobility } \newline glass\_mobility.mp4 \newline ($\mathcal{C} = 0.222$, $\phi=0.693$) \newline Mobility of the glass phase.
    \item \textbf{Isotropic fluid on a triangular surface} \newline triangular\_fluid.mp4 \newline ($\mathcal{C} = $, $\phi=0.7$) \newline Isotropic fluid on a triangular surface. 
    \item \textbf{Frustrated square solid on a triangular surface} \newline
    triangular\_frustratedSquareSolid.mp4 \newline ($\mathcal{C} = $, $\phi=0.75$) \newline Frustrated square solid on a triangular surface. 
    \item \textbf{Frustrated square solid-hexagonal coexistence} \newline
    triangular\_frustratedSquareSolid\_hexagonal.mp4 \newline ($\mathcal{C} = $, $\phi=0.8$) \newline Coexistence of the frustrated square solid and hexagonal phases on a triangular surface. 
    
\end{enumerate}

\section{Binodals}

\begin{figure}
    \centering
    \includegraphics[width=.45\textwidth]{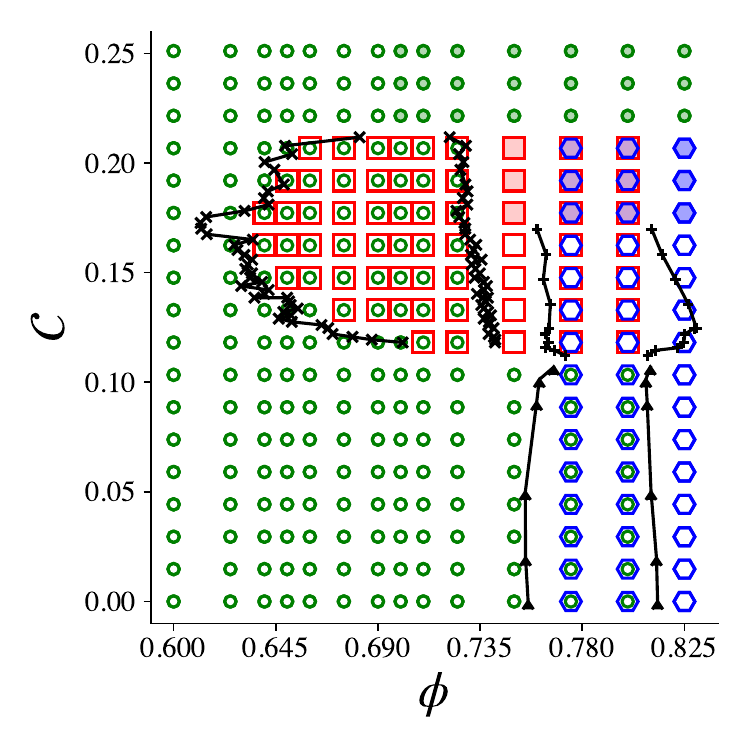}
    \caption{\protect\small{{Phase diagram of pentavalent patchy particles with coexistence binodals for square solid-isotropic fluid coexistence, square solid-hexagonal coexistence and hexagonal-isotropic fluid coexistence. Marker labels are provided in the main text.}}}
    \label{fig:PhaseDiagramBinodals}
\end{figure}

The three regions of coexistence described in the main text are square solid-fluid, hexagonal-fluid and hexagonal-square solid. 
The binodals defining these regions can be determined by extracting the phase densities as a function of curvature for several bulk surface coverages as described in the Methods section of the main text. 
The binodals for each coexistence region is displayed along with the full phase diagram in Fig.~\ref{fig:PhaseDiagramBinodals}.

At surface coverage $0.6 < \phi < 0.775$, increasing curvature to a minimum $\mathcal{C}=0.12$ results in square solid-fluid coexistence.
The square solid-fluid coexistence boundary shows strong curvature dependence. 
The fluid phase density (left binodal) first decreases with increasing curvature and then increases while the solid phase density (right binodal) monotonically decreases. 

At larger surface coverage $0.775 < \phi < 0.825$ low curvature assemblies feature hexagonal-fluid coexistence. 
The phase densities show small dependence on curvature, the fluid phase density slightly increases with increasing curvature and the hexagonal phase density slightly decreases.  
From here, increasing curvature beyond $\mathcal{C} = 0.12$ results in square solid-hexagonal coexistence. 
In this coexistence scenario, the square phase density monotonically decreases with increasing curvature while the hexagonal phase density first increases and then decreases.   
At these high surface coverages, dynamics slow considerably at curvatures $\mathcal{C}>0.17$ and preclude us from observing thermodynamic equilibrium (shaded symbols in Fig.~\ref{fig:PhaseDiagramBinodals}).
Although we observe signs of phase separation, we do not include the phase densities in the square solid-hexagonal binodal. 

\section{Curvature dependence of the Bond Number}

\begin{figure}
	\centering
	\includegraphics[width=0.45\textwidth]{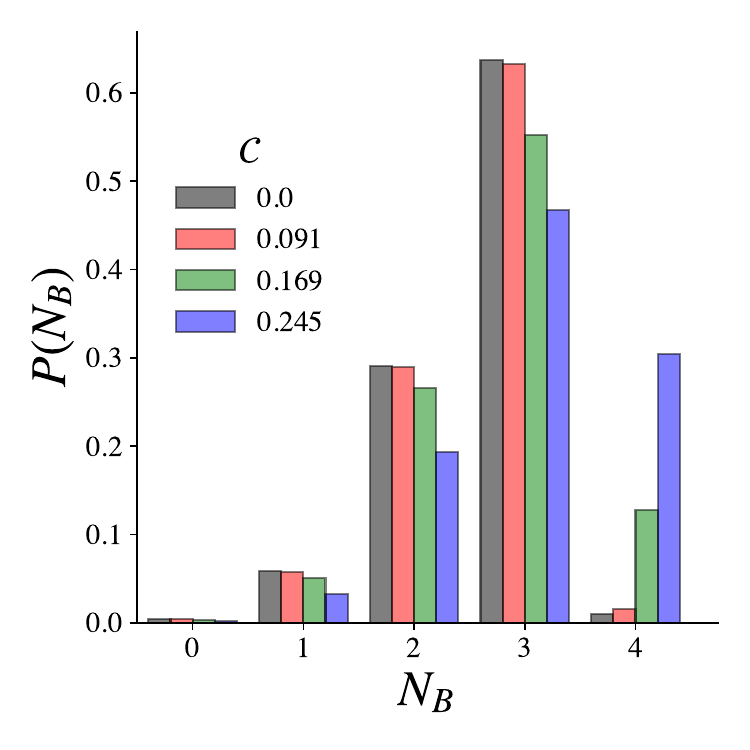}
	\caption{\protect\small{{Probability distributions of the number of bonds $N_B$ formed between a particle and its neighbors for several $\mathcal{C}$ and $\phi = 0.625$. }}}
	\label{fig:BondDistribution}
\end{figure}

The emergence and subsequent disappearance of square order with increasing surface curvature points to a nonmonotonic dependence of the relative change in free energy of the isotropic fluid and square solid as a function of $\mathcal{C}$.
As discussed in the main text, increasing $\mathcal{C}$ promotes bond formation in the isotropic fluid.
To illustrate this, we plot the probability distribution of the per particle bond number $P(N_B)$ for a isotropic fluid at a fixed surface coverage $\phi=0.625$ and several $\mathcal{C}$ in Fig.~\ref{fig:BondDistribution}.
For planar and low curvature $(\mathcal{C}< 0.1)$ few particles are able to form four bonds while, at higher curvatures, the peak at the maximum $N_B = 4$ rises significantly, diminishing the relative energetic benefit of square order. 
At sufficient $\mathcal{C}$, the entropically favorable isotropic fluid is the globally stable configuration for all surface coverages. 

\section{Bond orientational correlations}
In 2D systems, transitions between solid and fluid phases can be mediated by an intermediate $k$-atic phase, where $k$ is an integer value representing the type of bond orientational order present in the system. 
These states can be quantitatively distinguished from one another by the decay of the spatial bond orientational correlation function, $G_k(|\mathbf{r}|) = \langle \psi_k(\mathbf{r})\psi_k^*(\mathbf{0})\rangle$.
The decay of $G_k(|\mathbf{r}|)$  is exponential and algebraic for isotropic fluids and $k$-atic phases respectively. 
Solid phase bond orientational correlations are long ranged and do not decay.
Setting $k=4$ allows us to quantify tetratic ordering and to quantitatively distinguish the isotropic fluid, tetratic and square solid phases.
In our calculations $\mathbf{r}$ is the Euclidean displacement, however a more accurate analysis would use geodesic distances. 
We expect only a quantitative difference from our choice of Euclidean distance.

\begin{figure}
    \centering
    \includegraphics[width=.45\textwidth]{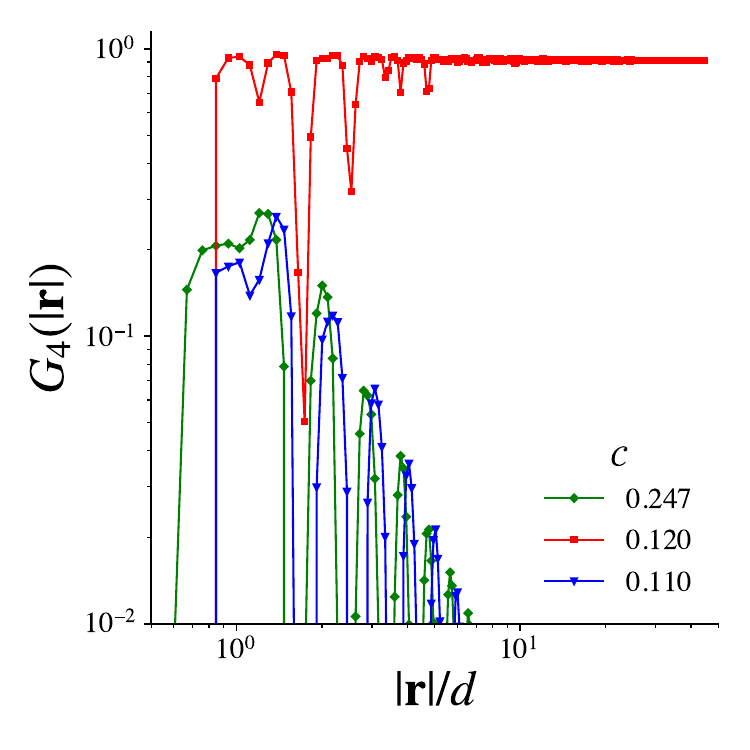}
    \caption{\protect\small{{Tetratic bond-orientational correlation functions for a fixed surface coverage $\phi=0.625$ of the isotropic fluid (blue), the square solid (red) and the dynamically arrested (green) state. The solid phase displays long-range tetratic order in contrast to the fluid states.}}}
    \label{fig:PureSolid}
\end{figure}

Figure.~\ref{fig:PureSolid} displays $G_4(|\mathbf{r}|)$ at a fixed density of $\phi=0.75$ and with varying $\mathcal{C}$. 
For curvatures below $(\mathcal{C} <0.10)$ and above $(\mathcal{C} >0.21)$, the assemblies show short-ranged correlations indicative of disordered phases.
Curvatures within the pure solid region $(0.10 < \mathcal{C} <0.17)$ show long-ranged correlations.
For $0.17 <\mathcal{C} < 0.21$ (shaded symbols Fig.~\ref{fig:PhaseDiagramBinodals}), the slow dynamics of the assembly preclude us from observing thermodynamic equilibrium, and we were unable to confirm the decay of correlations for the solid phase.  
One might expect that the transition between the isotropic fluid to the pure square phase may include a tetratic intermediate phase as observed in fluids of hard squares~\cite{Anderson2017ShapePolygons}.
However we did not observe any evidence of this in our simulations.

\begin{figure}
    \centering
    \includegraphics[width=.45\textwidth]{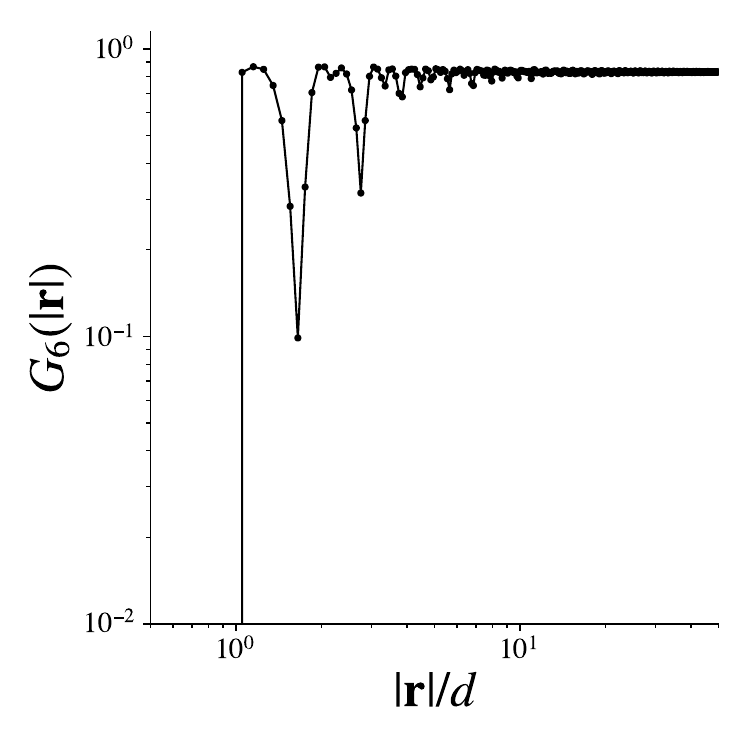}
    \caption{\protect\small{{Hexatic bond-orientational correlation function for $\phi=0.825$ and $\mathcal{C}=0$.}}}
    \label{fig:CFHex}
\end{figure}

From the hexagonal-isotropic fluid binodal, we determine the hexagonal phase density to be $\phi \approx 0.813$ for the planar surface.
As mentioned in the main text, to determine the nature of this phase we conduct a simulation at slightly higher density than $0.813$ in order to examine this phase when spatially uniform. 
The hexatic bond orientational correlation of a planar system at a surface coverage of $\phi=0.825$ is shown in Fig.~\ref{fig:CFHex}.
The absence of a discernible spatial decay is indicative of long-ranged orientational correlations and solid-like behavior.
However, the transition between hexatic to hexagonal solid is known to occur over a very small density range~\cite{Kapfer2015Two-DimensionalTransitions} so it remains possible that a hexatic to hexagonal solid transition occurs between $\phi = 0.813$ and $0.825$.
Locating and examining this transition (if present) as a function of surface curvature is the subject of future work. 

\begin{figure*}
	\centering
	\includegraphics[width=1\textwidth]{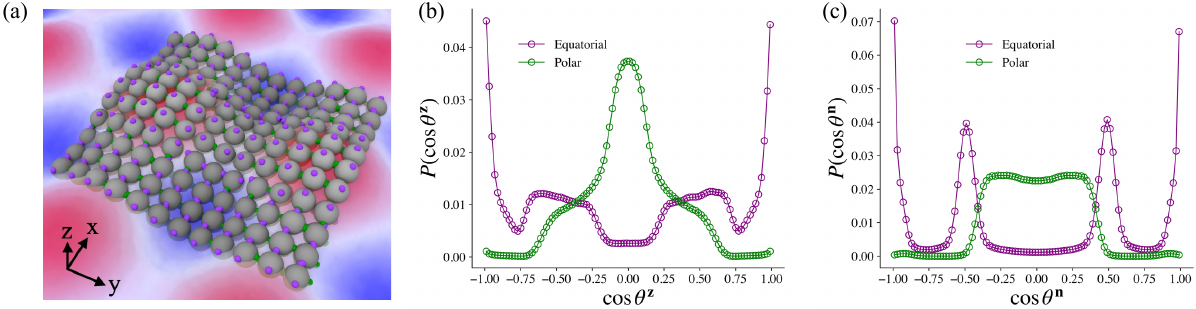}
	\caption{\protect\small{{Patch orientation and positioning on the square lattice. (a) Perspective snapshot of a small section of the square lattice. Equatorial patches are colored purple and polar patches are colored green. The underlying surface is colored by it's height difference from the flat plane, blue indicating valleys and red indicating peaks. Patches are enlarged for visual clarity. } Probability distribution for (b) $\theta^{\mathbf{z}}$ and (c) $\theta^{\mathbf{n}}$ at a fixed surface coverage $\phi=0.75$ and $\mathcal{C}= 0.12$ such that the system is in the pure square solid phase. }}
	\label{fig:SquareCrystalwithOrientations}
\end{figure*}

\section{Orientational alignment distributions of the square solid}
The emergence of square order from pentavalent patchy particles is intriguing as the additional binding site per particle, in addition to the misalignment of patches from the basal plane, would seem to preclude the formation of four-fold coordinated lattices. 
This is true for planar surfaces where we find no surface coverage that results in structures with square order (see Fig.~\ref{fig:PhaseDiagramBinodals}).
However, increasing the surface curvature in our system allows for interactions between patches off the $x-y$ plane. 
This serves to enable four-fold coordination and stabilize the square lattice.
As shown in Fig.~\ref{fig:SquareCrystalwithOrientations}(a), the patches of particles in a square solid are arranged such that on average the bond number is four.
Additionally, Fig.~\ref{fig:SquareCrystalwithOrientations}(a) appears to show a coupling between particle orientation and curvature. 
Local maxima of the surface (shown in red) and their connecting saddle regions are populated by particles with an equatorial patch (in purple) orientated upward, toward the positive $z$ direction, while particles residing in local minima (in blue) and their connecting saddle regions are oriented downward.

To quantify this perceived orientation-curvature coupling we compute distributions of particle orientations with respect to surface and axial directions. 
We define the patch orientation vector for patch $p$ on a patchy particle with position $\mathbf{r}$ as $\mathbf{q}_{p}= \mathbf{r}_p - \mathbf{r}$, where $\mathbf{r}_p$ is the position of the patch and $\mathbf{r}$ is the center position of the core particle.
We then compute the angle between the $z$-axis and the patch orientation vectors of each patchy particle, $\cos{\theta_{p}^{\mathbf{z}}}=\left(\hat{\mathbf{z}} \cdot \hat{\mathbf{q}}_{p}\right)$ as well as the angle with the surface normal $\cos{\theta_{p}^{\mathbf{n}}}=\left(\hat{\mathbf{n}} \cdot \hat{\mathbf{q}}_{p}\right)$, where $\hat{\mathbf{q}},\hat{\mathbf{z}},\hat{\mathbf{n}}$ are unit vectors.  
Examining the orientational distribution of $\cos{\theta^\mathbf{z}}$ and $\cos{\theta^\mathbf{n}}$ for each patch type makes clear their respective roles in bonding.

Patch orientation distributions for a pure square solid with density $\phi=0.75$ are shown in Fig.\ref{fig:SquareCrystalwithOrientations}(b) and (c). 
The distribution of $\cos\theta^\mathbf{z}$ for polar patches displays a sharp peak at $\cos\theta^\mathbf{z}= 0$, indicating that polar patches are predominately aligned perpendicular to the z-axis and parallel to the $x-y$ plane. 
The distribution for equatorial patches show peaks favoring alignment ($\cos\theta^\mathbf{z} = 1$) and anti-alignment  ($\cos\theta^\mathbf{z} = -1$) with the z-axis and two broad peaks of $\cos\theta^\mathbf{z}$ from $[-0.75 , -0.25] $ and $\cos\theta^\mathbf{z}$ from $[0.75 , 0.25] $.
The $\cos\theta^\mathbf{n}$ distribution has sharp peaks at $\cos\theta^\mathbf{n}= -1, -0.5, 0.5$ and $1$ for the equatorial patches, while the distribution for polar patches shows a broad peak of $\cos\theta^\mathbf{n}$ spanning $[-0.4 , 0.4] $.
We identify the angles with sharp peaks in these distributions as the patch orientations responsible for bonding and the formation of the square lattice.
From these sharp peaks, we observe that polar patches show strong correlation with the z-axis, while equatorial patches show strong correlation with the surface normal.
In the absence of curvature, $\hat{\mathbf{z}}= \hat{\mathbf{n}}$, preventing square solid formation in planar systems through the discussed alignment mechanism.

It can be shown through simple geometric arguments that a patchy particle with an equatorial patch aligned with the surface normal must also have two equatorial patches with $\cos \theta^\mathbf{n} = -0.5$. 
Likewise, a patchy particle with an equatorial patch anti-aligned must also have two equatorial patches with $\cos \theta^\mathbf{n} = 0.5$. 
We can thus simplify our representation of orientational texturing by focusing only on the non-redundant alignment and anti-alignment orientations, whose spatial distribution is shown in Fig.~4 in the main text.

For each patchy particle in the square solid, at least one patch must be excluded from bonding at any given time.
This additional patch may switch into a bonding orientation due to thermal fluctuations and/or particle rearrangement. 
Due to the indistinguishability between patches of a specific type, bond switching happens equally between patches.
This extra degree of freedom likely reduces the entropic penalty associated with square order.

\section{Orientational Alignment Distributions of the hexagonal phase}
In our system hexagonal order, unlike square order, is not induced by surface curvature.
We therefore do not expect to see a strong coupling between particle orientation and the local properties of the surface for particles in the hexagonal phase. 
To test this expectation we compute particle orientational alignment distributions of a pure hexagonal assembly on a planar surface and compare it to that of a curved surface at a fixed surface coverage of $\phi=0.825$. 

\begin{figure}
    \centering
    \includegraphics[width=0.4\textwidth]{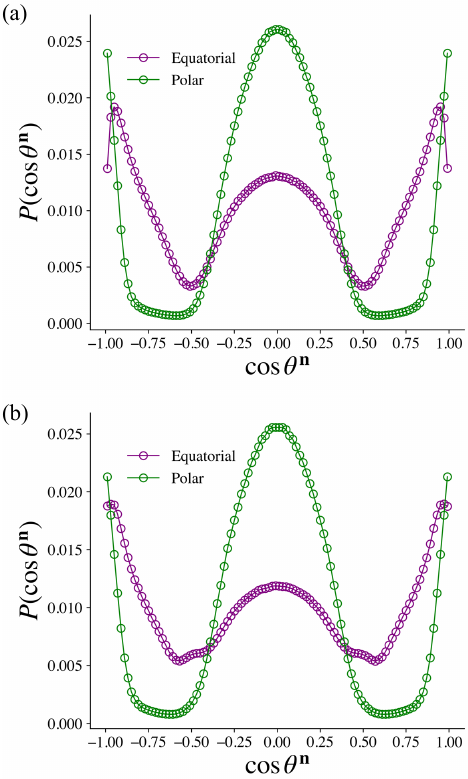}
    \caption{\protect\small{{ {Probability distribution of orientational alignment with the surface normal for a pure hexagonal assembly with surface coverage $\phi=0.825$ on (a) a planar surface and (b) on a surface with $\mathcal{C}=0.116$.} }}}
    \label{fig:HexDist}
\end{figure}

Figure~\ref{fig:HexDist}(a) shows distributions of patch alignment with the surface normal for both equatorial and polar patches on a planar surface. 
Both patches show a broad peak of $\cos\theta^\mathbf{n}$ from $[-0.50, 0.50]$ and peaks at $\cos\theta^\mathbf{n}=1$ and $\cos\theta^\mathbf{n}=-1$ indicating alignment and anti-alignment.
There is a notable absence of polar patches with alignment between $\cos\theta^\mathbf{n}=0.6$ and $\cos\theta^\mathbf{n}=0.8$ as well as between $\cos\theta^\mathbf{n}=-0.8$ and $\cos\theta^\mathbf{n}=-0.6$. 

Figure~\ref{fig:HexDist}(b) shows distributions of patch alignment with the surface normal for a surface with $\mathcal{C}=0.116$.
The distribution of polar patch alignment remains largely unchanged from the planar surface. 
The distribution for equatorial patches shows a small bump at alignment angles of $\cos\theta^\mathbf{n}=-0.5$ and $\cos\theta^\mathbf{n}=0.5$ that are indicative of the orientational texturing present in the square solid.
However, upon visualizing the spatial distribution of equatorial patch alignment we see no indication of orientational texturing at the scale visible in the square solid.
Since square order is not observed at any large scales for these assemblies we suspect these small peaks in the distribution are due to transient arrangements of local square order that are not globally stable at this surface coverage and curvature.
Furthermore, polar patches in the hexagonal phase do not exhibit any preferential alignment at scales that would indicate an obvious orientational texturing as the equatorial patches do within the square phase.
As the patch alignment distributions display only small quantitative changes between planar and curved surfaces, we can conclude that hexagonal order is not strongly dependent on the surface curvature. 

\begin{figure}
	\centering
	\includegraphics[width=0.45\textwidth]{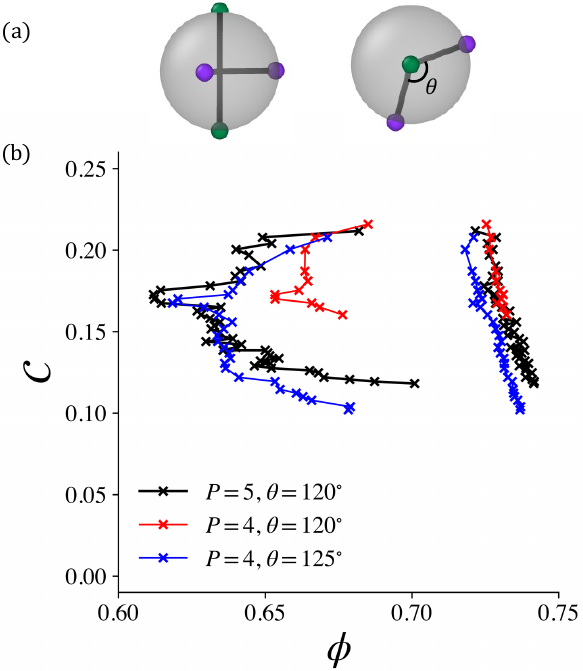}
	\caption{\protect\small{{Effect of particle geometry and patch number on the solid-isotropic fluid coexistence region. (a) Geometry of the ``seesaw'' particles considered with $\theta$ being the angle between equatorial patches. Equatorial patches are in purple and polar patches are in green. (b) Square solid-isotropic fluid coexistence binodals for pentavalent particles (black) and for seesaw particles with varying equatorial patch angle (red and blue). }}}
	\label{fig:GeometryBinodals}
\end{figure}

\section{Effect of particle geometry on coexistence}
The square solid phase emerges at a surface coverage dependent $\mathcal{C}$ that is intimately tied to the geometry of the patchy particle.
As discussed in the main text and the preceding section, the positioning of the equatorial patches is crucial for formation of the square solid. 
Adjusting the angle $\theta$ between the equatorial patches is thus expected to alter the coexisting boundaries and should then allow for phase separation at smaller values of $\mathcal{C}$.   

To isolate the effect of equatorial angle we focus on particles with four patches, arranged in a ``seesaw'' geometry such that there are two polar and two equatorial patches [Fig.~\ref{fig:GeometryBinodals}(a)]. 
We investigate two angles, $\theta = 120^{\circ} $ (which corresponds to the same equatorial arrangement as for the pentavalent particles) and $\theta = 125^{\circ}$. 
The binodals for the pentavalent particles and the two seesaw geometries are shown in Fig.~\ref{fig:GeometryBinodals}(b). 
For seesaw particles with $\theta = 120^{\circ}$ the coexistence region is smaller than for the pentavalent particles, and phase separation does not occur until exceeding $\mathcal{C} = 0.16$. 
This would seem to indicate that the non-bonded patch plays a more nuanced role in stabilizing the square solid.
As previously discussed, pentavalent particles are capable of switching their bonding equatorial patches while maintaining four bonds with their neighbors.
On the other hand, seesaw particles are forced to maintain the same orientation in order to maintain square order. 
Thus the square solid formed by pentavalent particles is likely more robust to thermal fluctuations, promoting the stability of the square solid at a broader range of $\mathcal{C}$ than for seesaw particles. 
Particles with $\theta = 125^{\circ}$ phase separate at curvatures as low as $\mathcal{C}=0.10$ and have a larger coexistence region then pentavalent particles. 
Thus larger angles can expand the curvatures over which square order is favorable. 
These findings suggest engineering particle anisotropy is a viable method toward tuning the region of coexistence and generating desired phase behavior.  

\section{Effect of surface symmetry on phase behavior}
\begin{figure}
	\centering
	\includegraphics[width=.4\textwidth]{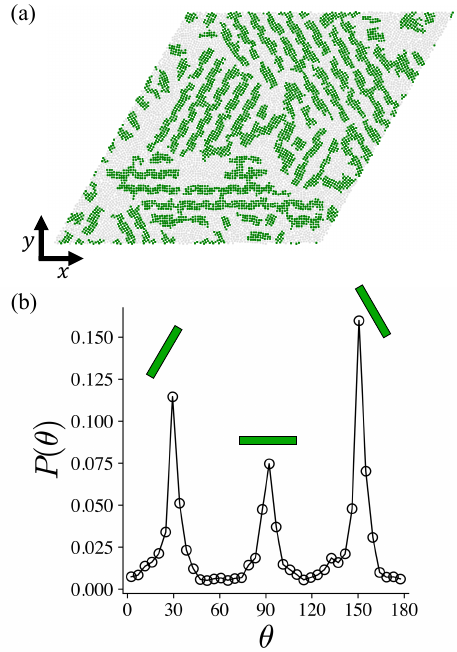}
	\caption{\protect\small{{{Frustrated square solid cluster orientation on a triangular surface. (a) Spatial distribution of particles with $|\psi_4|>0.9$ in green and $|\psi_4|<0.9$ in gray. (b) Distribution of cluster orientations with respect to the y-axis. The green bars indicate the orientation of each cluster corresponding to the three peaks. }}}}
	\label{fig:HexWaveCluster}
\end{figure}

In the main text, we discuss the effect of surface symmetry on phase behavior and provide a phase diagram of particle assembly on a triangular surface.
We find several notable differences from the phase diagram of assemblies on a square symmetric surface.
Most notably, the square solid no longer coarsens into one single domain. 
Instead, we observe a fragmented assembly of square solid clusters.
A representative snapshot of this assembly, which we term as a ``frustrated square solid,'' is shown in Fig.~\ref{fig:HexWaveCluster}(a). 
Particles with $|\psi_4|>0.9$ (in green) form long, thin, highly asymmetric clusters surrounded by particles with small $|\psi_4|$ (in gray).
The lengths of these clusters can span many periods of the surface.
Interestingly, it appears the long-axis of these clusters takes one of three orientations, corresponding to the edges of an equilateral triangle.

To quantify this observation, we compute the orientation of each cluster's long-axis by determining the principle direction corresponding to the largest principle moment for each cluster's radius of gyration tensor.  
The probability distribution of the angle ($\theta$) between the principle direction (long-axis) and the y-axis averaged over many clusters is shown in Fig~\ref{fig:HexWaveCluster}(b). 
The sharp peaks at $\theta = 30^{\circ}$, $90^{\circ}$ and $150^{\circ}$ indicate the preferred orientations of the clusters.
These orientations are spaced by $60^{\circ}$, which corresponds to the orientation of the edges of an equilateral triangle and the unit cell of the triangular surface.
At equilibrium, we speculate clusters should take a single orientation since clusters with dissimilar orientation cannot fuse together along their long axis into a single cluster (thereby reducing the systems free energy).
Which of the three orientations the system chooses is random.
The asymmetry of the probability distribution in Fig~\ref{fig:HexWaveCluster}(b) suggests that the system is not at equilibrium and is gradually relaxing toward favoring a cluster orientation of $\theta = 150^{\circ}$.

\begin{figure}
	\centering
	\includegraphics[width=.4\textwidth]{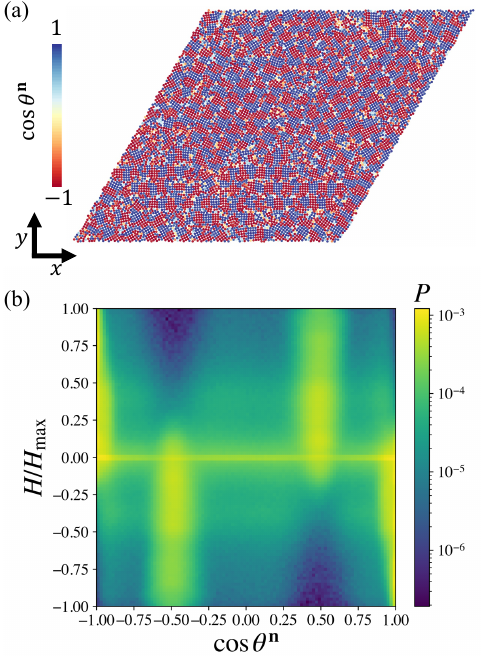}
	\caption{\protect\small{{  Orientational texturing of a frustrated square solid assembly on a triangular surface. (a) Spatial distribution of equatorial patch alignment with the surface unit normal $\hat{\mathbf{n}}$, colored by the most aligned or anti-aligned equatorial patch. Aligned particles are in blue while anti-aligned particles are in red. (b) Joint probability distribution of mean curvature $H$ and $\theta^\mathbf{n}$ for the frustrated square solid phase.}}}
	\label{fig:HexWaveNormal}
\end{figure}

Next we investigate whether the origin of square order on the triangular surface follows the orientational texturing mechanism present on a square surface.
The spatial distribution of the particle alignment with the unit surface normal $\hat{\mathbf{n}}$ is shown in Fig.~\ref{fig:HexWaveNormal}(a) and is colored the same as Fig.~4 in the main text.
Though in general there is more disorder, the assembly displays the same orientational texturing shown in the square symmetric surface with alternating regions of alignment and anti-alignment.
Indeed, upon computing the 2D probability distribution of mean curvature and particle orientation, we see that the same correlation between aligned/anti-aligned with negative/positive mean curvature persists as in the square symmetric surface [Fig.~\ref{fig:HexWaveNormal}(b)]. 
Thus we can conclude that the particle-scale origin of square order is consistent between the triangular and square symmetric surfaces.

\section{Curvature dependence of the chemical potential}
In a previous section, we discussed how increasing curvature enhances bond formation and thus provides an enthalpic contribution that reduces the free energy. 
In order to measure the effect of curvature on the free energy, we conduct simulations on a composite surface that features a curved region and planar region as discussed in the main text. 
The composite surface is defined as: 
\[
S(x,y) =
\begin{cases}
h\cos{(\frac{2\pi}{\lambda}x)}\cos{(\frac{2\pi}{\lambda}y)}, & |x| < L \text{ and } |y| < L \\
0, & |x| > L \text{ or } |y| > L
\end{cases}
\]
Where $L$ is a length chosen such that $S(x,y)$ smoothly transitions between curved and planar regions while maintaining an equal surface area $\mathcal{A}$, between the curved and planar regions.
The surface is shown in schematically in Fig.~8(a) in the main text.
For this choice of $S(x,y)$ there are two peak-to-peak distances: $\lambda$ and $\lambda' =  \sqrt{2}\lambda/2$. 
We find the smallest peak-to-peak distance is the most relevant length scale with which to define the curvature, thus $\mathcal{C} = d/\lambda'$.

We conduct simulations and monitor if particles migrate to or away from the curved region, reporting the equilibrium density difference between the two regions, $\Delta \phi = \phi_{\mathcal{C}}-\phi_{\mathcal{P}}$, where $\mathcal{C}$ denotes the curved region and $\mathcal{P}$ denotes the planar region.
A positive (negative) $\Delta \phi$ would indicate particles ``prefer'' the curved (planar) region.
Furthermore, the chemical potential must be spatially uniform at equilibrium. 
Thus the particles in the two regions must have the same chemical potential. 
We propose the chemical potential is a function of both $\phi$ and $\mathcal{C}$. 
Equality of chemical potential then results in 
\begin{equation}
\mu(\phi_{\mathcal{C}},\mathcal{C})=\mu(\phi_{\mathcal{P}},0).
\label{eq:DeltaMu}
\end{equation} 
We can expand the chemical potential in the curved region to first order in $\Delta \phi$ and $\mathcal{C}$,
\begin{equation}
\mu(\phi_{\mathcal{C}},\mathcal{C})=\mu(\phi_{\mathcal{P}},0) + \frac{\partial \mu}{\partial \mathcal{\phi_{\mathcal{C}}}}\Bigr|_{ \phi_{\mathcal{C}}=\phi_{\mathcal{P}}}\Delta \phi + \frac{\partial \mu}{\partial \mathcal{\mathcal{C}}}\Bigr|_{\mathcal{C}=0}\mathcal{C}.
\end{equation}
Using Eq.~\eqref{eq:DeltaMu}, we find
\begin{equation}
\frac{\partial \mu}{\partial \mathcal{\phi}}\Bigr|_{ \phi=\phi_{\mathcal{P}}}\Delta \phi = -\frac{\partial \mu}{\partial \mathcal{\mathcal{C}}}\Bigr|_{\mathcal{C}=0}\mathcal{C}.
\end{equation}
Since thermodynamic stability requires $\partial \mu/\partial \phi > 0$, a $\Delta \phi >0$ suggests that $\partial \mu/\partial 
\mathcal{C}<0$ for small $\mathcal{C}$.
Thus through this methodology, we can indirectly probe the leading order curvature dependence of the chemical potential for small curvature.

We can make a theoretical prediction for $\Delta \phi$ as a function of curvature for hard-sphere and ideal particles.
We consider our exact simulation setup: particles in three dimensions experiencing a strong harmonic trap centered at the surface, $S(x,y)$.
We emphasize that the particles have full 3D translational and rotational degrees of freedom but are strongly bound to the surface such that quantities, such as the surface coverage $\phi$, remain well defined.
The canonical partition function of $N$ particles confined to the surface takes the form:
\begin{equation}
Q =\frac{1}{N!\Lambda^{3N}}\int{\mathrm{d}\mathbf{r}^N\exp{\left[-\beta\left( u^{\mathrm{int}} + u^{\mathrm{surf}}\right)\right]}},
\end{equation}
where $u^{\mathrm{surf}}$ is the confining surface potential and $u^{\mathrm{int}}$ is the particle interaction potential.
Here, $\Lambda$ is our spatial resolution (the de Broglie wavelength for molecular systems). 

In general, particle interactions will prevent the exact analytical evaluation of the partition function. 
Our aim here, however, is to identify the form of the chemical potential of the particles and to introduce suitable approximations for estimating the terms that emerge.
In the limit of strong confinement, we expect the assembly to effectively be confined to our 2D surface, and we thus expect to recover the 2D ``ideal gas'' contribution to the chemical potential at low surface coverage. 
With this in mind, we can introduce the following useful definitions 
\begin{subequations}
\begin{equation}
    Q = Q^{\rm ideal}Q^{\rm ex},
\end{equation}
where $Q^{\rm ideal}$ is the canonical partition function of a 2D ideal collection of particles:
\begin{equation}
    Q^{\rm ideal} = \frac{\mathcal{A}^N}{N!\Lambda^{2N}}, 
\end{equation}
and $Q^{\rm ex}$ represents the factor of the partition function in ``excess'' to the ideal contribution with:
\begin{equation}
    Q^{\rm{ex}} = \frac{1}{(\mathcal{A}\Lambda)^N}\int\mathrm{d}\mathbf{r}^N\exp\left[-\beta\left( u^{\mathrm{int}} + u^{\mathrm{surf}}\right)\right].
\end{equation}
\end{subequations}
The free energy of the system takes the form ${F = -k_BT\ln Q = F^{\rm ideal} + F^{\rm ex}}$ where ${\beta F^{\rm ideal} = -\ln Q^{\rm ideal} = N\left(\ln(\rho\Lambda^2)-1\right)}$ and ${\beta F^{\rm ex} = -\ln Q^{\rm ex}}$.
Here, $\rho = N/\mathcal{A}$ is the 2D surface number density. 
The chemical potential then follows as $\mu = \partial F/\partial N$ with:
\begin{subequations}
\begin{equation}
\mu(\phi, \mathcal{C}) = \mu^{\rm ideal}(\phi) + \mu^{\rm ex}(\phi, \mathcal{C})
\end{equation}
where the ideal chemical potential takes the familiar form:
\begin{equation}
\mu^{\mathrm{ideal}}=k_BT\ln{(\Lambda^2 \rho)},
\end{equation}
and the excess portion of the chemical potential is:
\begin{equation}
\mu^{\mathrm{ex}}= \frac{\partial}{\partial N} F^{\rm ex}.
\end{equation}
\end{subequations}
We emphasize that the excess chemical potential generally depends on the surface coverage $\phi$ (through the interaction potential) and the surface \textit{curvature} $\mathcal{C}$ through the confining surface potential. 

Before attempting to build a theory for $\mu^{\mathrm{ex}}$ it is instructive to consider the non-interacting limit.
This limit can be recovered by either identically setting $u^{\rm int} = 0$ or more physically as $\phi \rightarrow 0$.
In the absence of interactions, the $N$ spatial integrals in the excess partition function become decoupled and it can then be expressed as ${Q^{\rm ex} = (\mathcal{J}/\Lambda)^N}$ where we have defined a length scale associated with the surface trap, ${\mathcal{J}= \int{\mathrm{d}\mathbf{r}\exp{\left(-\beta u^{\mathrm{surf}}\right)}}/\mathcal{A}}$.
In this limit, the excess chemical potential then takes the form ${\lim_{\phi\rightarrow0} \mu^{ex} = k_BT \ln (\Lambda/\mathcal{J})}$. 
Generally, the chemical potential is thus not identically that of a strictly two-dimensional assembly, even in the ideal limit.
We will denote this limiting value of the excess chemical as ${\mu^{\rm surf}(\mathcal{C}) \equiv k_BT \ln (\Lambda/\mathcal{J})}$ which only depends on the surface curvature.

It is perhaps conceptually more convenient to construct a theory for the excess chemical potential for a \textit{strictly} two-dimensional assembly on a curved surface $S(x,y)$ without considering the precise physics that binds the particle to the surface. 
This 2D excess chemical potential, $\mu^{\rm ex}_{\rm 2D}$, would then vanish in the limit of vanishing surface coverage, much like a traditional excess chemical potential. 
We can then approximate our original, quasi-2D excess chemical potential as ${\mu^{\rm ex}(\phi, \mathcal{C}) \approx \mu^{\rm ex}_{\rm 2D}(\phi, \mathcal{C}) + \mu^{\rm surf}(\mathcal{C})}$.

With our approximate form of the chemical potential:
\begin{equation}
\mu(\phi, \mathcal{C})=\mu^\mathrm{ideal}(\phi) + \mu^{\mathrm{surf}}(\mathcal{C})+\mu_{\rm 2D}^\mathrm{ex}(\phi, \mathcal{C}),
\label{eq:chemicalPotential}
\end{equation}
all that we need is a theory for $\mu^{\rm ex}_{\rm 2D}$.
Here, we will limit our analysis to hard disks which have a well-established thermodynamics literature. 
For hard-disks, we can estimate $\mu^{\rm ex}_{\rm 2D}$ by using a scaled particle theory (SPT) equation of state derived for hard-disks on curved surfaces.
Before doing so, we first derive the equation of state ($\mu^{\rm ex}_{\rm 2D}$) on a planar surface and will then extend this result to surfaces with non-uniform curvature.
This derivation follows that of Refs.~\cite{Lishchuk2009, Hansen2013TheoryEdition}. 

The excess chemical potential of a fluid of hard disks is equivalent to the reversible work required to insert a single hard-disk into the fluid.
We first consider the work required to create a cavity of radius $r_0$, $W(r_0)$. 
The equilibrium probability of observing a cavity of size $r_0$ within the fluid follows as:
\begin{equation}
    p_0(r_0) \propto \exp{\left[-\beta W(r_0)\right]}.
    \label{eq:Probability0}
\end{equation}
As the overlap of hard-disks is forbidden, there can be at most a single particle whose center lies in a cavity with radius ${r_0 < r}$.
The probability of finding a particle within a cavity of this size is:
\begin{equation}
    p_1(r_0) = \rho a(r_0) = 1 - p_0(r_0),
    \label{eq:Probability1}
\end{equation}
where $a(r)$ is the surface area of a circle with radius $r$.
On a planar surface, $a(r) = \pi r^2$.
The combination of Eqs.~\eqref{eq:Probability0} and~\eqref{eq:Probability1} allows us to identify the reversible work for creating a small cavity (${r_0 < r}$) as:
\begin{subequations}
\begin{equation}
    W(r_0) = -k_BT\ln{\left[ 1 - \rho a(r_0 ) \right]}.
    \label{eq:W_2}
\end{equation}
For large cavities $(r_0 \gg r)$, the reversible work can be expressed using the bulk thermodynamic pressure (i.e.,~we consider the ``$P-V$'' work for create the cavity):
\begin{equation}
    W(r_0) = P a(r_0),
    \label{eq:revWorkLargeCavity}
\end{equation}
where $P$ is the pressure of the fluid. 

We now have a form of the reversible work for cavity creation in two-limiting cases. 
To determine the work as a continuous function of $r_0 \ge r$ between these limits we propose the following expansion:
\begin{equation}
    W(r_0) = w_0 + w_1 \left(r_0 - r \right) + P a\left(r_0 - r\right),
    \label{eq:W_1}
\end{equation}
where $a$ is evaluated at a radius of ${r_0-r}$. 
We can appreciate that the final term in the above expansion is the quadratic term and will dominate for $r \gg r_0$, recovering the anticipated reversible work from bulk thermodynamics.
The coefficients $w_0$ and $w_1$ can be found by requiring the work $W(r_0)$ and its derivative $\partial W/\partial r_0$ given by Eqs.~\eqref{eq:W_2} for ${r_0 < r}$ and~\eqref{eq:W_1} for ${r_0>r}$ to be continuous at ${r_0 = r}$. 
Therefore, the coefficients are:
\begin{equation}
\begin{aligned}
    \beta w_0 &= -\ln{\left( 1- \phi \right)}, \\
    \beta w_1 &= \frac{ 2\pi r \rho }{1-\phi},
\end{aligned}
\end{equation}
where we have defined the area fraction of particles as $ {\phi = \rho \pi r^2}$.
The reversible work for $r_0 \geq r$ is then
\begin{equation}
    \beta W(r_0) = -\ln{\left( 1- \phi \right)} +  \frac{ 2 \pi r\rho }{1-\phi} (r_0 -r ) + P\pi(r_0 -r)^2
    \label{eq:Work}
\end{equation}
\end{subequations}

The excess chemical potential is equivalent to the reversible work required to insert a hard-disk of radius $r$ into the fluid. 
Since the minimum separation distance between the centers of neighboring hard-disk particles is $2r$, the work required to insert a disk of radius $r$ is equivalent to the work required to create a circular cavity with radius $2r$ that excludes the centers of all other particles. 
Thus we can evaluate Eq.~\eqref{eq:Work} with $r_0=2r$ to find $\mu^{\rm{ex}}_{2D}$ on a planar surface:
\begin{equation}
\begin{aligned}
\beta \mu_{\rm 2D}^\mathrm{ex}(\phi, \mathcal{C}=0) = -\ln{\left[1 - \phi\right]} + \frac{2\phi}{1-\phi}   + \frac{\beta P \phi}{\rho}.
\end{aligned}
\label{eq:mu_planar_excess}
\end{equation}
We can eliminate the pressure from our expression by considering the additional condition offered by the Gibbs-Duhem relation:
\begin{equation}
    \frac{\partial P}{\partial \rho} = \rho \frac{\partial \mu}{\partial \rho}
    \label{eq:thermo_relation}
\end{equation}
where $\mu$ is the total chemical potential. 
The final expression for the planar excess chemical potential follows as:
\begin{equation}
\beta \mu_{\rm 2D}^\mathrm{ex}(\phi, \mathcal{C}=0) = -\ln{\left[1 - \phi\right]} + \frac{2\phi}{1-\phi}   + \frac{\phi}{\left(1-\phi\right)^2}.
\label{eq:mu_planar_excess_final}
\end{equation}

On a curved surface one must take into account the effect of the local Gaussian curvature $K$ on $a(r)$. 
Following Ref.~\cite{Lishchuk2009}, we take $a(r)$ to be the area of a geodesic disk on a two dimensional Riemann manifold (valid for small Gaussian curvature, $K \ll 1/r^2$):
\begin{equation}
    a(r) = \pi r^2 \left(1 - \frac{Kr^2}{12} \right) + \mathcal{O}(r^6).
\end{equation}
As the surface curvature in our system varies spatially, we consider the spatially-averaged Gaussian curvature with, ${K = \iint{\mathrm{d}x\mathrm{d}y K(x,y)\sqrt{1+(\nabla S)^2}}/\mathcal{A}}$.
The coefficients in the reversible work are then:
\begin{equation}
\begin{aligned}
    \beta w_0 &= -\ln{\left[ 1- \rho a(r) \right]} \\
    \beta w_1 &= \frac{4 \rho a(r) - 2 \phi}{r\left[1-\rho a(r)\right]}.
\end{aligned}
\end{equation}
We follow the same arguments presented for the planar surface to derive the excess chemical potential on a \textit{weakly} curved surface:
\begin{equation}
\begin{aligned}
\beta \mu_{\rm 2D}^\mathrm{ex}(\phi, \mathcal{C}) = & -\ln{\left[1 - \rho a(r)\right]} + \frac{4\rho a(r) - 2\phi}{1 - \rho a(r)} \\ 
& + \rho a(r)\frac{1-\phi +\rho a(r)}{[1 - \rho a(r)]^2},
\end{aligned}
\end{equation}
where we have again used the Gibbs-Duhem relation [Eq.~\eqref{eq:thermo_relation}].

\begin{figure}
	\centering
	\includegraphics[width=0.45\textwidth]{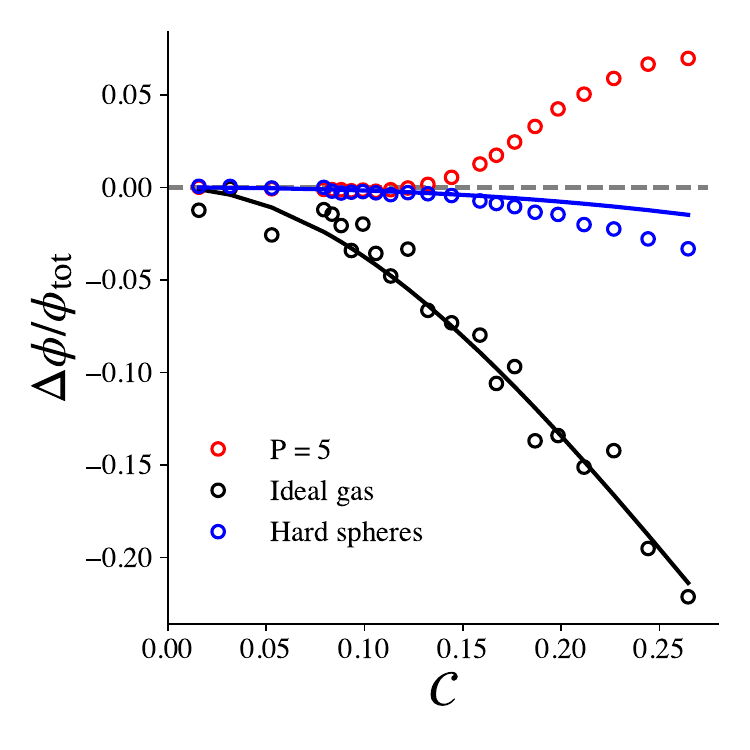}
	\caption{\protect\small{{ The difference in surface coverage between curved and planar regions $\Delta \phi = \left(\phi_\mathcal{C}-\phi_\mathcal{P}\right)$ as a function of $\mathcal{C}$ at a fixed total surface coverage of $\phi_\mathrm{tot}=0.6$ for pentavalent patchy particles (red) hard sphere particles (blue) and non-interacting ideal particles (black). The solid blue and black lines are the predictions from our theory.} }}
	\label{fig:DeltaPhi}
\end{figure}

We now have a theory for the chemical potential of ideal particles and hard disks on curved surfaces.
We can apply these expressions to model the chemical potential of hard-\textit{spheres} using Eq.~\eqref{eq:chemicalPotential}.
In the expressions for $\mu^{\rm ideal}$ and $\mu^{\rm surf}$ we set $\Lambda^2 = \pi r^2$. 
The chemical potential, equilibrium condition [Eq.~\eqref{eq:DeltaMu}], and constraint on the total surface coverage ($\phi_{\mathcal{C}}+ \phi_{\mathcal{P}} = 2\phi_{\mathrm{tot}}$ as the curved and planar regions occupy equal area) allow us to predict $\Delta \phi$ as a function of $\mathcal{C}$.
The relative difference $\Delta \phi/\phi_{\mathrm{tot}}$ as a function of $\mathcal{C}$ is shown in Fig.~\ref{fig:DeltaPhi} for an assembly of pentavalent particles (red), hard-spheres (blue) and an ideal gas (black).
We fix the total surface coverage at $\phi_{\mathrm{tot}} = 0.6$ so that the assembly is in the isotropic fluid phase for the full range of $\mathcal{C}$.
The estimate from our model for ideal and hard-sphere particles are the solid blue and black lines, respectively.

Interestingly, $\Delta \phi$ for ideal particles shows strong curvature dependence. 
This is entirely due to $\mu^{\rm surf}$ and the quasi-2D nature of our particles.
As $\mathcal{C}$ increases the boundaries of the curved region on the $x$-$y$ plane must shrink so that the surface area of the curved region remains equivalent to the planar region.
This means that the 3D volume available to particles in the curved region is reduced relative to particles in the planar region. 
The result is a bias for $\Delta \phi$ to be increasingly negative with increasing $\mathcal{C}$, even in the absence of particle interactions.
This effect would be absent for particles living on a strictly 2D surface.
However, real systems of particle assemblies on surfaces are only quasi-2D. 
Like the particles in our simulations, these assemblies live in 3D but are restricted translationally through some form of surface confinement. 
The exact role of this effect in experiments remains unclear.
While the bias induced by $\mu^{\rm surf}$ is a result of our simulation methodology, we have fully accounted for its effect in our interpretation of the results.

Hard-sphere particles show a monotonically decreasing $\Delta \phi$ as a function of $\mathcal{C}$. 
This is a direct result of the increased excluded area per particle $a(r)$ as a result of non-zero $K$. 
This entropic penalty increases the chemical potential of particles in the curved region relative to the planar region, driving $\Delta \phi$ to become negative.
At large curvature $(\mathcal{C}>0.15)$ our model fails to quantitatively emulate the simulation result because our expression for $a(r)$ is valid only for weakly curved surfaces.
Patchy particles are also subject to the same entropic penalties due to curvature. 
However, increasing surface curvature allows for additional patch-patch interactions between neighboring particles.
These interactions provide an enthalpic benefit that reduces the chemical potential of particles within the curved region. 
For $0 < \mathcal{C} < 0.11$ pentavalent patchy particles follow the result of hard-spheres and the prediction from SPT.
However, for $\mathcal{C} > 0.11$ enthalpic benefits overcome entropic penalties and particle density in the curved region increases, resulting in a positive $\Delta \phi$ with increasing $\mathcal{C}$ [red markers in Fig.~\ref{fig:DeltaPhi}]. 
This analysis provides a preliminary understanding of how surface curvature impacts the free energy landscape of surface assemblies. 
Both entropic and enthalpic contributions to the free energy are affected in non-trivial ways by surface curvature.